\documentclass[conference]{IEEEtran}
\IEEEoverridecommandlockouts
\usepackage{cite}
\usepackage[T1]{fontenc}
\usepackage{color}
\usepackage{amsmath,amssymb,amsfonts}
\usepackage{algorithmic}
\usepackage{graphicx}
\usepackage{textcomp}
\usepackage{array} 
\renewcommand\arraystretch{1.1}
\usepackage{url}
\usepackage{xcolor}
\usepackage{booktabs}
\usepackage{makecell}
\makeatletter
\renewcommand{\maketag@@@}[1]{\hbox{\m@th\normalsize\normalfont#1}}%
\makeatother
\usepackage{algorithm}
\usepackage{algorithmic}
\usepackage{array}
\usepackage{multirow}
\usepackage{amsmath}
\usepackage{graphics}
\usepackage{epsfig}
\usepackage{amssymb}
\usepackage{url}
\usepackage{makecell}
\usepackage{amsmath}
\usepackage{subfigure}
\usepackage{balance}
\usepackage{tabularx}

\usepackage{stackengine}

\usepackage{caption}
\newfloat{figtab}{htb}{fgtb}
\makeatletter
\newcommand\figcaption{\def\@captype{figure}\caption}
\newcommand\tabcaption{\def\@captype{table}\caption}% default is 1.0
\makeatother

\begin{document}
	
	\title{
		
		From Ember to Blaze: Swift Interactive Video Adaptation via Meta-Reinforcement Learning
		
		%Enhancing Large-Scale Real-World Interactive Video Adaptation via Meta-Reinforcement Learning
		%Scale Semantics-Aware Mobile Vision with DNN-Driven Compressive Offloading
		%DNN-Driven Compressive Offloading: Bringing Semantic Video Segmentation to Mobiles
		%DNN-Driven Compressive Offloading for Edge-Assisted Semantic Video Segmentation 
		%Spatiotemporal Adaptive Compression for Edge-Assisted Semantic Segmentation	
	}	
	\author{\IEEEauthorblockN{Xuedou Xiao{${^\dag}{^\ddag}$}, Mingxuan Yan{${^\dag}{^\ddag}$}, Yingying Zuo{${^\dag}$}, Boxi Liu{${^\S}$}, Paul Ruan{${^\S}$}, Yang Cao{${^\dag}$}, Wei Wang{${^\dag}{^\ast}$}}\IEEEauthorblockA{{${^\dag}$}School of Electronic Information and Communications, Huazhong University of Science and Technology, China\\ {${^\S}$}Tencent Technology Co. Ltd, China\\ Email: \{xuedouxiao, mingxuanyan, yingyingzuo, ycao, weiwangw\}@hust.edu.cn, \{percyliu, paulruan\}@tencent.com}
	\thanks{${^\ast}$The corresponding author is Wei Wang (weiwangw@hust.edu.cn).} 
	\thanks{${^\ddag}$Both authors have equal contribution.} 
	}	
	\maketitle
	
	\begin{abstract}
		Maximizing quality of experience (QoE) for interactive video streaming has been a long-standing challenge, as its delay-sensitive nature makes it more vulnerable to bandwidth fluctuations. While reinforcement learning (RL) has demonstrated great potential, existing works are either limited by fixed models or require enormous data/time for online adaptation, which struggle to fit time-varying and diverse network states. Driven by these practical concerns, we perform large-scale measurements on WeChat for Business's interactive video service to study real-world network fluctuations. Surprisingly, our analysis shows that, compared to time-varying network metrics, network sequences exhibit noticeable short-term continuity, sufficient for few-shot learning requirement. We thus propose Fiammetta, the first meta-RL-based bitrate adaptation algorithm for interactive video streaming. Building on the short-term continuity, Fiammetta accumulates learning experiences through offline meta-training and enables fast online adaptation to changing network states through few gradient updates. Moreover, Fiammetta innovatively incorporates a probing mechanism for real-time monitoring of network states, and proposes an adaptive meta-testing mechanism for seamless adaptation. We implement Fiammetta on a testbed whose end-to-end network follows the real-world WeChat for Business traces. The results show that Fiammetta outperforms prior algorithms significantly, improving video bitrate by 3.6\%-16.2\% without increasing stalling rate.

	\end{abstract}
	
	\begin{IEEEkeywords}
		Interactive video streaming, bitrate adaptation, meta-reinforcement learning
	\end{IEEEkeywords}
	
	\section{Introduction}
 	Recent years have witnessed the evolution of video streaming from traditional video on demand (VoD), live TV to ultra-low-latency interactive video applications, such as WeChat, Skype, Zoom, Facetime, etc. Especially with the  outbreak of COVID-19 that bounds people with social distancing, the demand for digital classrooms~\cite{classroom}, video conferences~\cite{conference}, e-commerce~\cite{ecommerce}, etc. has increased substantially. Polaris Market reports that the global interactive video market is expected to reach a staggering \$10.23 billion by 2028~\cite{market}. %Looking into the near future, more new applications are rapidly emerging, from cloud gaming~\cite{gaming}, VR streaming~\cite{VR} to telesurgery~\cite{telesurgery}. %Interactive video has de facto reshaped our lives.

 	%Aside from that, with the step-by-step upgrade of mobile communication devices (LTE-Advanced, 5G, WiFi6, the envisioned 6G), more new interactive video use cases are rapidly emerging, ranging from cloud gaming, VR streaming to telesurgery. To this day, real-time interactive video communication 
 	
 	%dominates today's Internet traffic.

 	Despite the fast-paced development, the quality of experience (QoE) of interactive video streaming remains unsatisfactory, such as annoying ultra-blurry images and frequent stalling. The intrinsic reason is that interactive video streaming is highly susceptible to bandwidth fluctuations, due to \textit{(\romannumeral1)} the strictest latency requirement (e.g., 200~ms) that limits the buffer and resilience to bandwidth fluctuations; \textit{(\romannumeral2)} the RTP/UDP protocol that hardly achieves reliable transmission; \textit{(\romannumeral3)} the instant capture and encoding that are more likely to waste bandwidth, compared to VoD like bulk data transfer.

	To improve the interactive video QoE, extensive research effort has been devoted along bitrate adaptation algorithms. Yet, whether rule-based~\cite{carlucci2016analysis,cardwell2017bbr,varma2015internet} or learning-based algorithms~\cite{mao2017neural,zuo2022adaptive,yan2020learning,zhou2019learning,zhang2020onrl,zhang2021loki,zhang2022deepcc,ma2022multi,huang2020quality} have their own limitations. \textit{(\romannumeral1)} Rule-based algorithms~\cite{carlucci2016analysis,cardwell2017bbr,varma2015internet} commonly adopt universal pre-programmed rules. Much recent literature has shown that these fixed rules can hardly fit diverse and time-varying network states caused by heterogeneous networks (e.g., WiFi, 4G, 5G), complex conditions (e.g., mobility, indoor/outdoor, density), etc. For example, Google Congestion Control (GCC)~\cite{carlucci2016analysis}, implemented in WebRTC, encounters overly conservative policies and a lack of agility, which lead to severe bandwidth wastage. \textit{(\romannumeral2)} Existing offline-learned neural networks (NNs)~\cite{mao2017neural,zuo2022adaptive,yan2020learning,zhou2019learning} are also constrained by their fixed parameters, which fall into the same dilemma as rule-based algorithms.
	\textit{(\romannumeral3)} Existing online learning-based bitrate adaptation algorithms~\cite{zhang2020onrl,zhang2021loki,zhang2022deepcc,ma2022multi,huang2020quality} (mainly transfer learning), however, require a large amount of data/time in the face of changing network states, making it difficult to achieve fast adaptation. Moreover, the real-time updates of NNs may converge to local optimum and the trial and error further degrades performance.

	\begin{figure*}[t]
		\begin{minipage}{0.33\linewidth}
			\centering
			\includegraphics[width=0.9\linewidth]{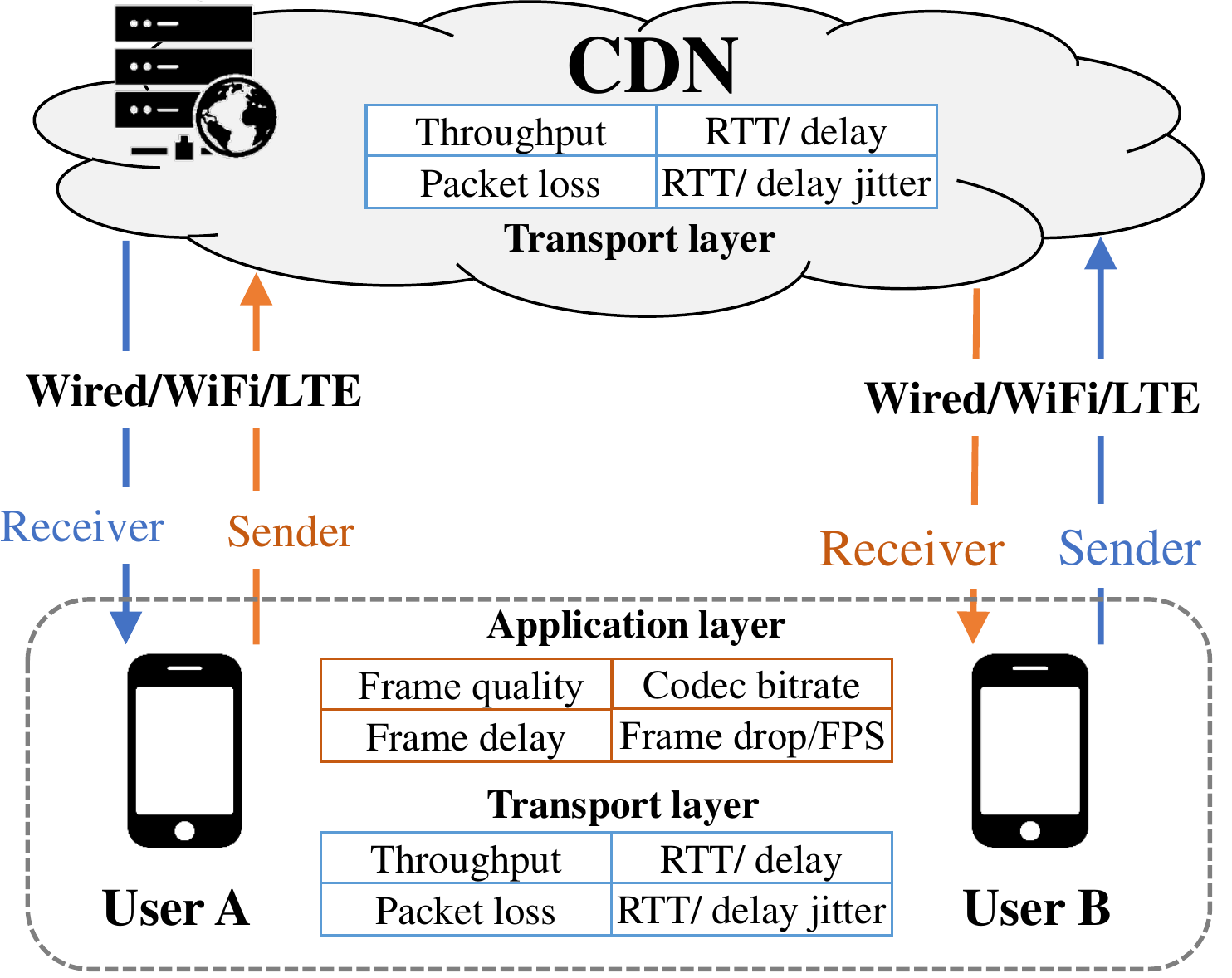}
			\caption{Interactive video system architecture of WeChat for Business.}
			\label{fig1}
		\end{minipage}\enspace
		\begin{minipage}{0.32\linewidth}
			\centering
			\includegraphics[width=0.9\linewidth]{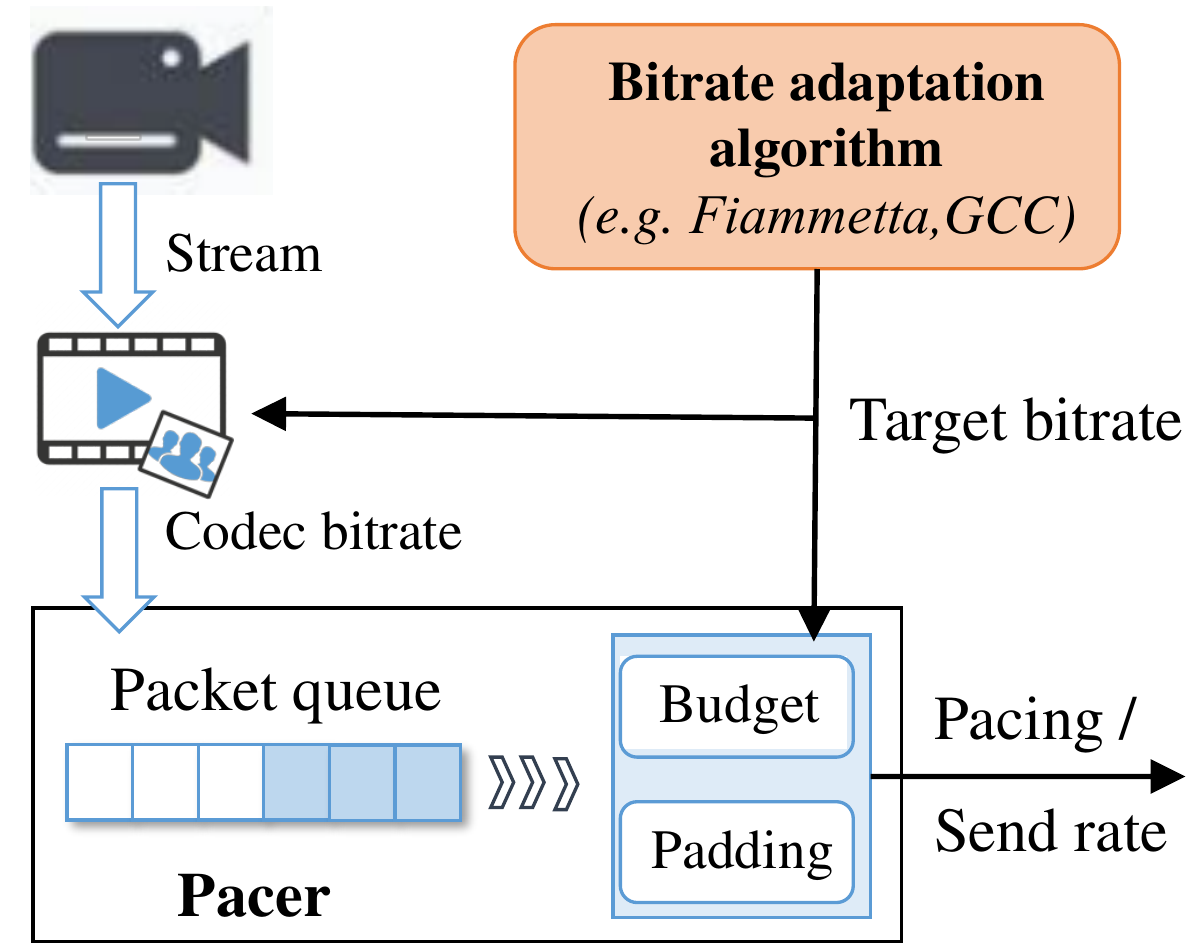}
			\figcaption{Built-in bitrate control module of interactive video system.}\label{fig2}
		\end{minipage}\enspace
		\begin{minipage}{0.33\textwidth}
			\centering
			\begin{tabular}{|m{0.35\textwidth}<{}||m{0.47\textwidth}<{\centering}|}
				\hline
				\textbf{ Features} &\textbf{Descriptions} \\
				\hline \hline
				Time span &  2022.1.13-15, 2022.5.15-17\\ 
				\hline
				Video time & 1402382~s\\
				\hline
				Video sessions  &  14428 \\
				\hline
				Network types &  Wired, WiFi, LTE (4G, 5G) \\
				\hline
				Countries & Over 200 countries\\
				\hline
				Volume & 7~Gb \\
				\hline
			\end{tabular}
			\captionof{table}{Dataset descriptions.}
			\label{table1}
		\end{minipage}\vspace{-0.3cm}

	\end{figure*} 
	
	Driven by these practical concerns, we seek to answer a key question: \textit{Can interactive video achieve fast adaptation to diverse and time-varying network states, pushing QoE to the limit? } To this end, we first conduct large-scale measurements of real-world network fluctuations on WeChat for Business interactive video service. The analysis shows that despite dramatic fluctuations in network metrics, there exists noticeable short-term continuity in network sequences defined by $ \{$mean\,$\mu$, std dev\,$\sigma$, fluctuation\,$\omega$, ranges\,$\boldsymbol{\Delta}\}$, which is sufficient for few-shot learning requirement. %In other words, the few-shot learning speed can be even faster than network state changes.
	
	Inspired by the short-term continuity, we propose Fiammetta, the first meta-reinforcement learning (RL)-based bitrate adaptation algorithm for interactive video systems, aiming at maximizing QoE. Fiammetta builds on the merit of ``learning to learn'' by retaining past learning experiences (i.e., offline meta-training), so as to adapt quickly online with a handful of gradient updates (i.e., meta-testing) to changing network states. Meta-learning is a typical framework for few-sample settings, but Fiammetta is not merely a straightforward environment shift. Instead, it entails three unique challenges.

	%Driven by these practical concerns, we seek to answer a key question: \textit{Can interactive video system realize real-time specialization and continual adaptation to diverse, dynamic and unknown networks, pushing QoE to the limit?} To this end, we propose Fiammetta, the first meta-reinforcement learning (RL) based bitrate adaptation algorithm for interactive video system. 
	%Fiammetta builds on the merit of ``learning to learn'' by retaining past learning experiences (i.e., offline meta train), so as to adapt quickly online with a handful of gradient updates (i.e., meta test) to current changing network states. The key observation is that the NN specialization can be even faster than the network state changes, draw from a measurement study of Tencent's interactive video telephony. Specifically, the network may appear to fluctuate dramatically, but if we define different network states by weighing 
	%$ \{$Mean\,$\mu$, Std Dev\,$\sigma$, Fluctuation\,$\omega$, Ranges\,$\boldsymbol{\Delta}\}$ of the bandwidth, there exits a short-term continuity of network states sufficient for seamless adaptation.

	%Fiammetta is not merely a straightforward environment shift of meta-RL in new field of interactive video streaming. Instead, it entails three unique design challenges.
	
	\textit{(\romannumeral1) How to define ``tasks'' in interactive video streaming?} Meta-learning exploits learning experience from previous ``tasks'' to accelerate online learning. As a matter of course, we consider the adaptation to different ``network states'' as ``tasks''. The problem evolves to how to define ``network states'' for objectivity of ``tasks'', since a mess of network metrics such as loss, RTT, throughput, etc. are deeply affected by bitrate adaptation policies. Once theses metrics are used, the ``task'' cannot keep constant during policy learning. For this reason, we specifically adopt bandwidth and propagation delay, which characterize environment, traffic flow density, motion state, link length, etc., but marginally affected by bitrate selection. On top of that, we define $\{\mu$, $\sigma$, $\omega \}$ of the bandwidth sequence and set ranges $\boldsymbol{\Delta}$ for a ``task'' to enhance its continuity. 
	
	%Besides, we extract $ \{$ $\mu$, $\sigma$, $\omega$,  $\boldsymbol{\Delta}\}$ of the bandwidth sequence to make the network state definition include fluctuation properties and ranges for generation of long-term bitrate adaptation policies.
	\textit{(\romannumeral2) How to realize the real-time estimation of ``network state'' to determine if it is a new ``task''?}  %the network indicators that can be measured are throughput, packet loss ratio and latency. 
	Accurately estimating bandwidth poses challenges, since it cannot be directly measured like RTT, throughput, loss, etc. %that is common in interactive video streaming.
	Rule-based algorithms %use either throughput at past congestion state or the maximum throughput over a window as current bandwidth. This 
	exploit stable bandwidth probing mechanisms that increase bitrates additively/multiplicatively. %, and then update bandwidth estimation by throughput. 
	Existing RL methods leverage the risky trial and error to probe bandwidth.
	Meta-RL, however, is more likely to underestimate bandwidth when it increases, as sub-NN is more specialized to one/last ``task'', leading to fewer trial-and-error actions.  %Meta-RL, however, has lower trial-and-error probabilities, as each sub-NN is more specialized for one task relevance. %, but less generalized to others. This deficiency causes meta-RL to a lag in response to bandwidth growth, even worse, leading to flow starvation. 
	To tackle this problem, we innovatively integrate probing into meta-RL by proposing a bandwidth estimation and filtering mechanism. Specifically, the loss and RTT are used to filter the throughput sequence, which provides hints on network utilization. 
	%During the full-use stage, the throughput is treated as bandwidth, while 
	During the under-use stage, the bandwidth is estimated to be the throughput plus a probing value.
	 %and the range $\boldsymbol{\Delta}$ of ``network state'' is used to mitigate the effect of estimation bias.%. %with reference to BBR. 
	This design enables a quick estimation of ``network state'' and uses the ranges to mitigate the effects of bias.

	\textit{(\romannumeral3) How to guarantee seamless adaptation?} The time spent on meta-testing might cause a lag in adapting to the changing ``network states'', which in turn leads to performance degradation. To handle this issue, we propose an adaptive meta-testing mechanism that sets an activation threshold to $\frac{\boldsymbol{\Delta}}{2}$ of current ``network state'', instead of waiting for the detected attributes completely out of ranges $\boldsymbol{\Delta}$. Moreover, we configure proper $\boldsymbol{\Delta}$ to guarantee seamless adaptation according to measurements on WeChat for Business interactive video system.
	
	\textbf{Results.} We implement Fiammetta on a testbed whose end-to-end network follows real-world WeChat for Business traces. We also deploy 3 state-of-the-art solutions: rule-based GCC~\cite{carlucci2016analysis}, learning-based OnRL~\cite{zhang2020onrl}, and hybrid Loki~\cite{zhang2021loki}, using 389 hours of video sessions for a half-month evaluation. Compared with baselines, Fiammetta improves video bitrate by  3.6\%-16.2\% and cuts stalling rate by 6.3\%-21.9\%. %in application layer. Meanwhile, Fiammetta leads to 3.6\%-16.2\% larger throughput and decreases RTT by 1.1\%-7.4\% in transport layer.
		
	\textbf{Contributions.}  \textit{(\romannumeral1)} Through large-scale measurements, we dive into real-world network fluctuations and short-term continuity of network sequences. \textit{(\romannumeral2)} We propose Fiammetta, which to our knowledge is the first to deploy meta-RL for fast adaptation to changing network states and maximize QoE. \textit{(\romannumeral3)} We implement Fiammetta on a testbed whose end-to-end network follows the real-world WeChat for Business traces, and validate its superiority over state-of-the-art solutions.
	
	The remainder of this paper is organized as follows. \S\ref{finding} introduces the measurement study and short-term continuity of network sequences. \S\ref{design} elaborates on the system design. Implementation
	and evaluation are detailed in \S\ref{sec-implementation} and \S\ref{evaluation}. \S\ref{related} gives a literature review, followed by conclusion in \S\ref{conclusion}.

	 %the long-tail performance issues of modern learning-based algorithms for real-time video adaptation

	%We conduct measurement studies comprehensively analyze the limitations of current ABR strategies
	
	%the impact of diverse, rapidly fluctuating or unseen network conditions on current ABR strategies. Second, we introduce a live video adaptation algorithm based on MetaR. Third, experiments

	%We offload online learning process (i.e., meta test) to the server to speed it up, and we end up with a time duration of 2s. Therefore, it is necessary to guarantee a seamless transition between sub-NNs during meta-test stage, i.e., no performance degradation.
	
	\section{Measurements and Findings}\label{finding}
	In this section, we conduct a measurement study on WeChat for Business interactive video platform, to investigate the fluctuation characteristics of real-world network states.
	
	\subsection{Measurement Platform}
	
	\begin{figure*}[ht]
		\centering
		\subfigure[$\hat{B}$ value before and after $\Delta t$.]{
			\includegraphics[width=0.237\linewidth]{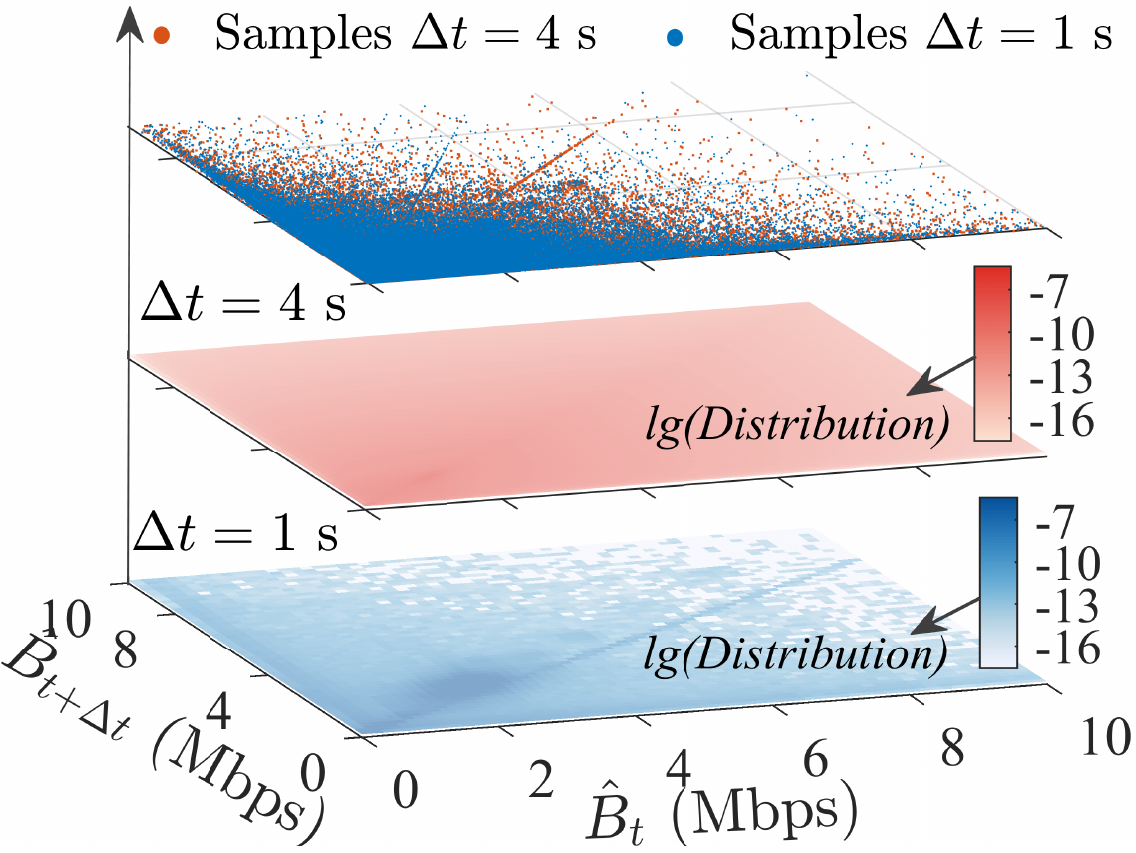}\label{fig3a}}
		\hfill
		\subfigure[$\mu$ value before and after $\Delta t$.]{
			\includegraphics[width=0.237\linewidth]{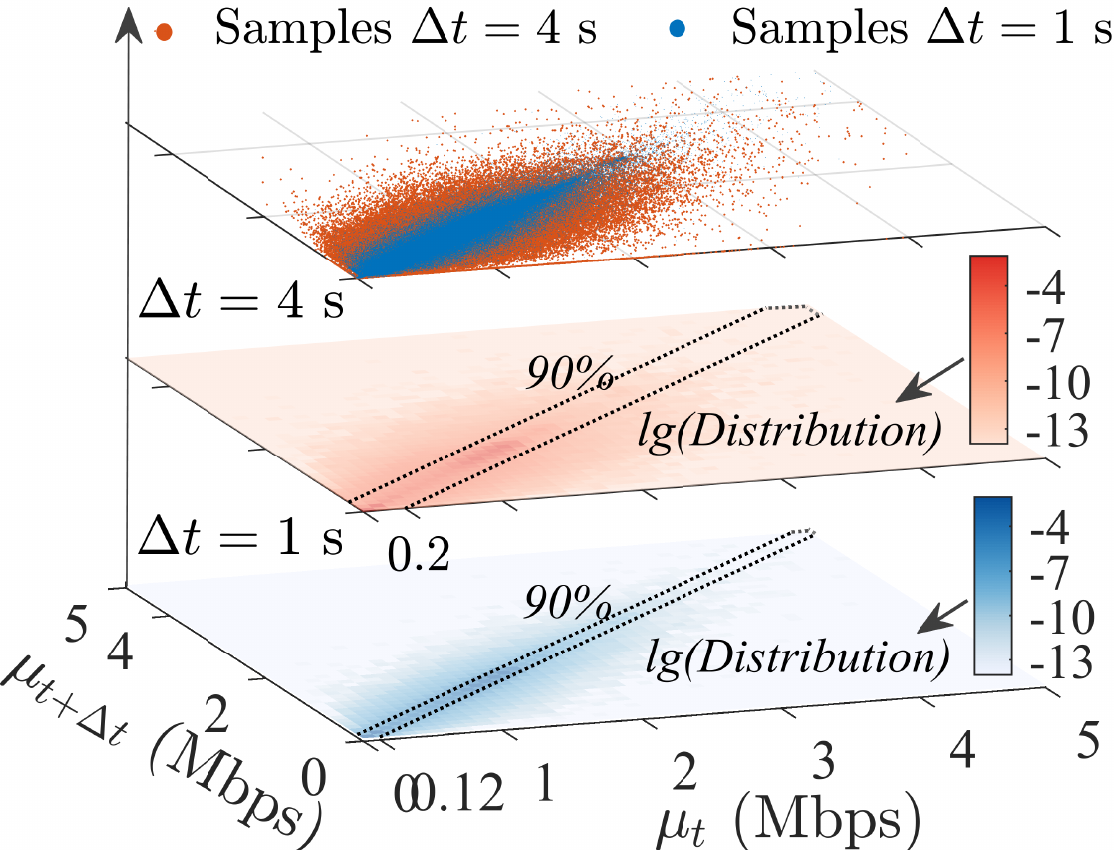}\label{fig3b}}
		\hfill
		\subfigure[$\sigma$ value before and after $\Delta t$.]{
			\includegraphics[width=0.237\linewidth]{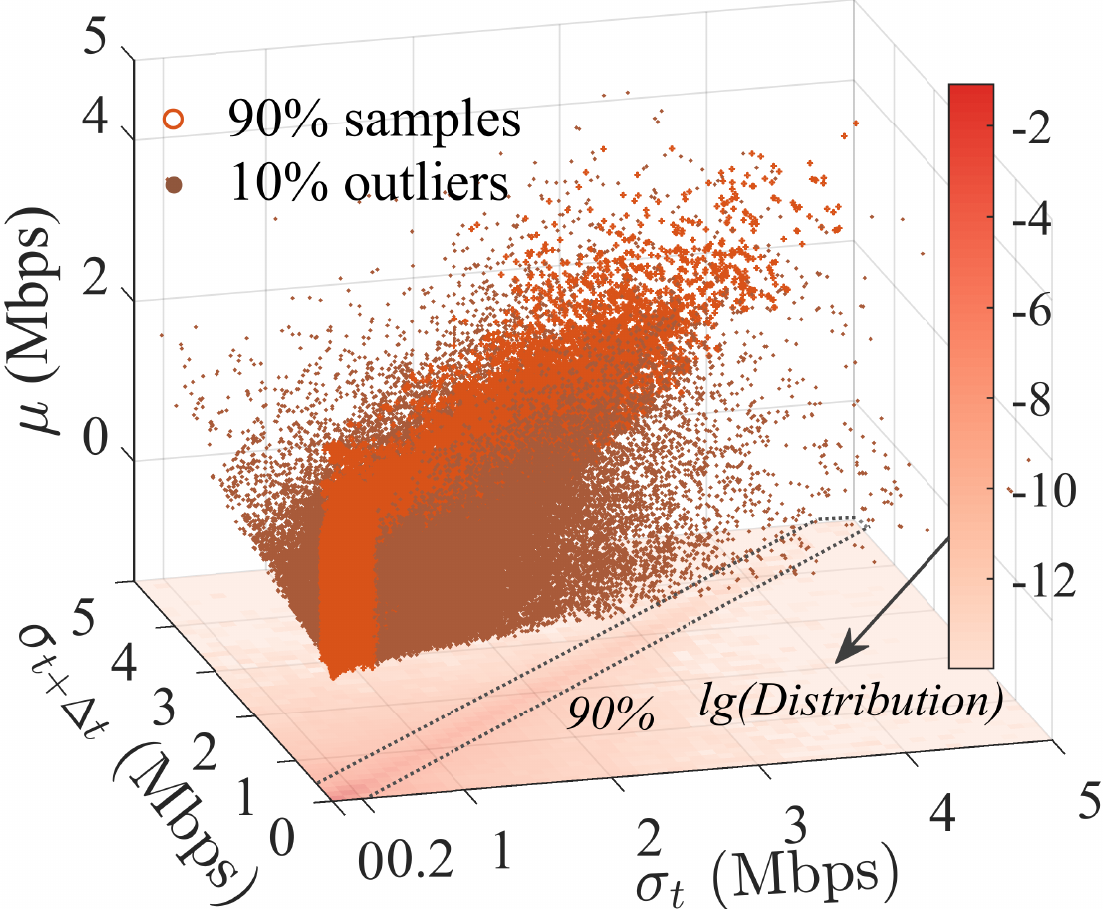}\label{fig3c}}
		\hfill
		\subfigure[$\omega$ value before and after $\Delta t$.]{
			\includegraphics[width=0.237\linewidth]{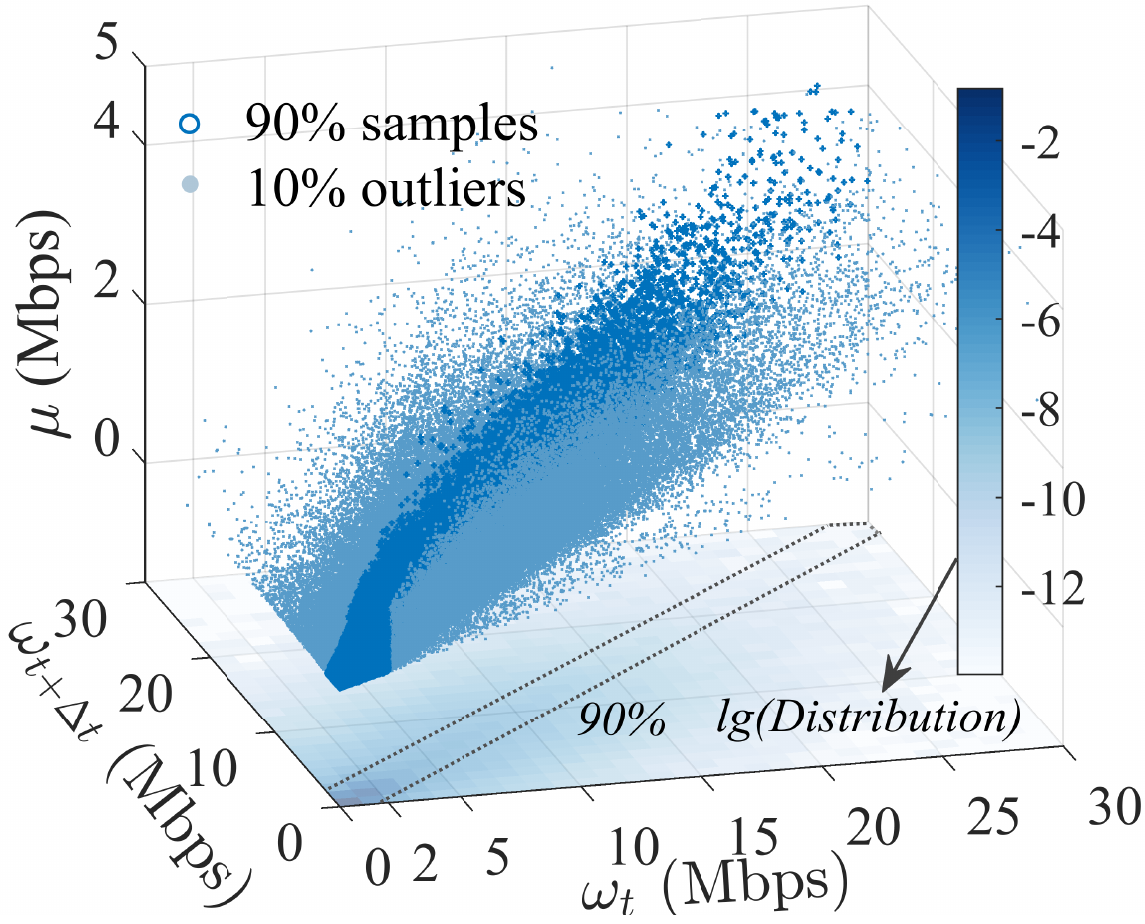}\label{fig3d}}
		\\
		\caption{Real-world network fluctuations.}\vspace{-0.2cm}
		\label{fig3} 
	\end{figure*}
	
	We log fine-grained end-to-end network metrics and video metrics on WeChat for Business's interactive video service worldwide. Fig.~\ref{fig1} and Fig.~\ref{fig2} depict the system architecture and the built-in bitrate control module. %Without exception, WeChat interactive video system is built on WebRTC, the de facto
	%video communication framework adopted by mainstream real-time video/audio applications such as hangout, Facebook, etc. 
	%As depicted in Fig.~\ref{fig2}, 
	Integrating transport and application layers, this system applies target bitrates estimated by the bitrate control module to both sending and codec bitrate adaptation.
	We establish logging points at the CDN and user sides. Therein, only users hold the video encoding and playback functions to execute both sending and codec bitrate adaptation, while the CDN is merely responsible for forwarding and sending bitrate adaptation in the transport layer. Therefore, we log both transport-layer metrics (i.e., throughput, RTT, RTT jitter, packet loss) and application-layer metrics (i.e., FPS, quantization parameter (QP), stalling rate and codec bitrate) at the user side, while only transport-layer metrics of the CDN are recorded. %It is noteworthy that due to the excessive amount of data, only transport layer metrics are recorded with 1~s granularity for network analysis and dataset collection. In contrast, during training and testing stage, the transport layer metrics are collected at 100~ms granularity to be fed into the neural network, and the application layer metrics are logged based on real-time video characteristics. 

	%It is noteworthy that all transport layer information, as well as target and codec bitrate are measured in 100ms timescale and the others logged in 1s timescale.

	\subsection{Dataset Descriptions} 
	Table~\ref{table1} summarizes the detailed information of our measurement dataset. Based on WeChat for Business APP, we collect network metrics corresponding to video sessions worldwide during two time spans of Jan 13-15 and May 15-17 with 1~s granularity. The entire dataset consists of 14428 video sessions, with an overall duration of over 1.4 million seconds and a total volume of 7~Gb. These video sessions are built on different heterogeneous networks (e.g., 4G, 5G, WiFi, wired), diverse user devices (e.g., cell phones, tablets, laptops, desktops and even smart watches), covering users in more than 200 countries worldwide. This dataset faithfully logs the real-network network fluctuations worldwide, under the influence of competing traffic, user movements, communication environments, network service providers and frequencies, and has broad coverage in both time and space.

	%span a period of time from X to Y

	%\subsection{Limitations of Existing Algorithms}
	
	\subsection{Short-Term Continuity of Network States}\label{continuity}
	Based on the dataset, we qualitatively analyze and quantitatively test the fluctuating characteristics of the network traces.
	
	As mentioned above, the dataset consists of throughput, loss and RTT. However, all these metrics are deeply affected by subjective factors (e.g., bitrate selections), %and objective factors (e.g., competing traffic, user movements), 
	which cannot unambiguously represent background network fluctuations (e.g., competing traffic, user movements). The best is available bandwidth, but it's hard to measure in real time at fine grain. %In fact, a plethora of research has uses throughput as available bandwidth, but most of them are dedicated to VoD. The adaptive bitrate (ABR) algorithms for VoD are only responsible for codec bitrate adaptation under throughput traces, without integrating sending rate (e.g., congestion control) to fit bandwidth and maximize throughput. 
	To handle this issue, we propose a novel bandwidth filtering and estimation mechanism that exploits measurable metrics to estimate bandwidth $\hat{B}$, detailed in \S\ref{preprocessing}, where we have verified its effectiveness. In what follows, we uses the $\hat{B}$ to investigate the network fluctuating characteristics.
	
	\textbf{Time-varying characteristic.} For a better visualization of network fluctuations, we group all estimated bandwidth pairs, with each pair consisting of estimated $\hat{B}$ before and after a time interval $\Delta t$ within the same video session. Fig.~\ref{fig3a} plots these pairs with $\Delta t$ of 1~s and 4~s in the form of scatter plots and further depicts their distributions. %, where the x-axis/y-axis represent $\hat{B}$ before/after $\Delta t$, respectively. 
	Specifically, we can notice the high diversity in the back-and-forth $\hat{B}$ fluctuations even at $\Delta t = 1$~s. The $\hat{B}$ value exhibits time-varying characteristics that even become more significant as $\Delta t$ increases. When $\Delta t$ reaches 4~s, the fluctuation tends to be more irregular and unpredictable, as the scatter points near the diagonal become less concentrated. In this case, the meta-testing (online adaptation) cannot catch up with the time-varying network fluctuations, when simply using bandwidth value as the definition of the ``network state''.

	\textbf{Short-term continuity.}	We turn to test whether the bandwidth sequence has predictable and regular fluctuating characteristics.  Similarly, we depict in Fig.~\ref{fig3b}-\ref{fig3d} the mean $\mu$, standard deviation $\sigma$, and fluctuation $\omega$ of the $\hat{B}$ sequences obtained from a sliding window $W_r$ before and after a time interval $\Delta t = 4$~s (related to the meta-testing time detailed in \S\ref{meta-testing}). Therein, $\omega$ is the sum of absolute adjacent value differences within $W_r$.
	It is obvious that these sequence properties exhibit more continuity. Even $\Delta t$ reaches 4~s, all the scatter points corresponding to the change in $\mu$, $\sigma$, $\omega$ converge at the diagonal. Furthermore, we find that 90\% of the points can be covered by extra adding a small range to $\mu$, $\sigma$, $\omega$. That is to say, if we define the ``network state'' of a learning task $i$ as a cluster of $\hat{B}$ sequence properties centered at $\{$$\mu_i$, $\sigma_i$, $\omega_i$$\}$ with covering ranges $\boldsymbol{\Delta}_i$, the ``network state'' presents a short-term continuity characteristics. 
	
	Since even few-shot learning (i.e., meta-testing process) takes a short period of time, the crux becomes how to define ``network state'' to gain more continuity and achieve seamless adaptation. A basic rule needs to be satisfied: for most time, the network fluctuations fall within the covering range of ``network state'' defined for the last ``task''. To alleviate the impact of meta-testing lag, the meta-testing process needs to be basically completed before the network changes beyond the range of the last ``task''. In short, the meta-testing needs to be faster than the ``network state'' changes. In this section, we observe the short-term continuity in network sequences, making it a reality to develop meta-RL-based video adaptation algorithm.

 	\section{System Design}\label{design}

 	\subsection{Overview}

\begin{figure}[t]
	\centering
	\includegraphics[width=0.98\linewidth]{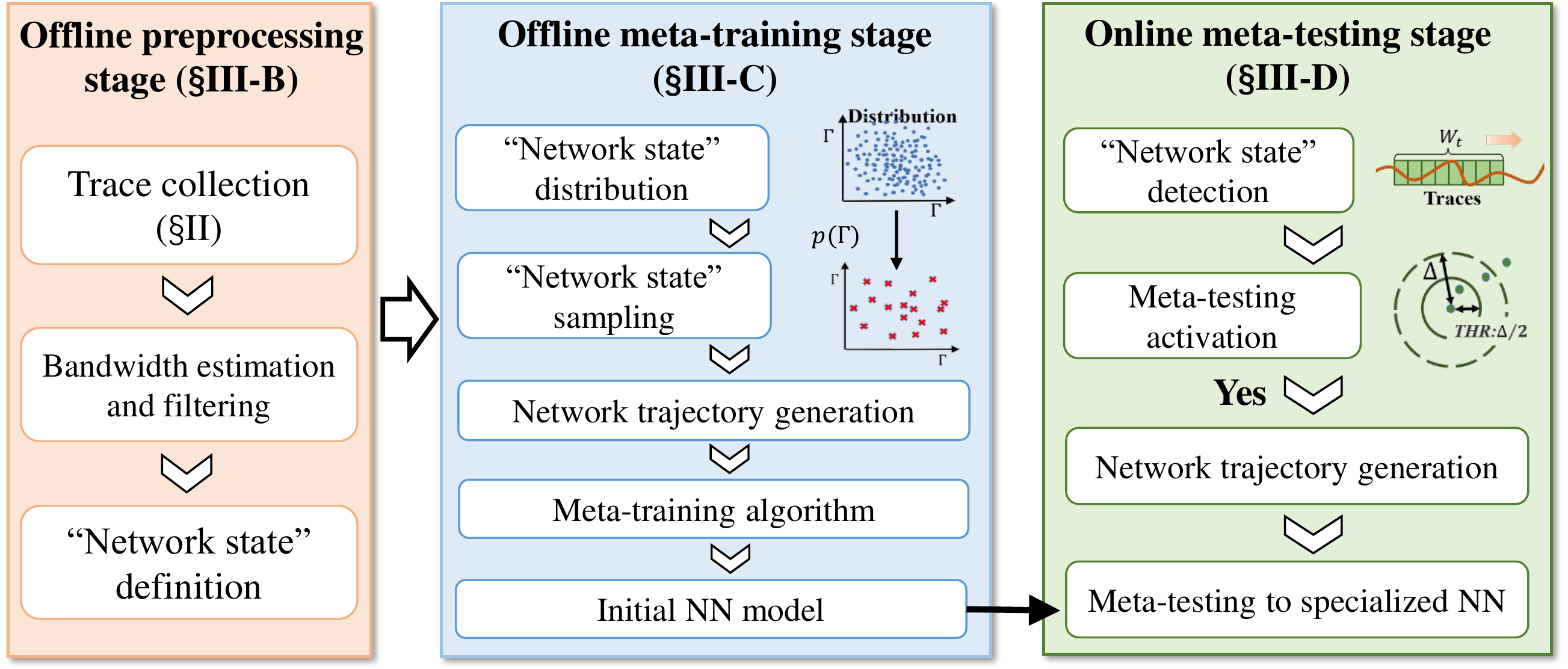}
	\caption{System overview.}\vspace{-0.3cm}
	\label{fig4}
\end{figure}

 	Fig.~\ref{fig4} shows the high-level overview of Fiammetta's design which contains three stages: pre-processing (\S\ref{preprocessing}), offline meta-training (\S\ref{meta-train}) and online meta-testing (\S\ref{meta-testing}). 
 	\begin{itemize}
 		\item [\textit{(\romannumeral1)}] 
 		\textit{Pre-processing} is the basis for subsequent meta-training and meta-testing due to its key steps of trace collection, bandwidth estimation and filtering, and the design and definition of ``network state''.    
 		\item [\textit{(\romannumeral2)}]
 		\textit{Meta-training} is implemented offline. The end goal is to obtain an initial NN model that can achieve fast adaptation and maximize QoE during meta-testing.
 		\item [\textit{(\romannumeral3)}]
 		\textit{Meta-testing} is performed online to seamless adapt to the changing ``network state'' based on real-time detection and meta-testing activation mechanism.

 		%\item [\textit{(\romannumeral4)}]
 		%\textit{Meta-training reactivation} is proposed  to maintain the adaptability of the initial NN model by configuring a threshold value of ``network state'' distribution changes.

 	\end{itemize}

 	\subsection{Pre-Processing Stage} \label{preprocessing}
 	
 	As shown in Fig.~\ref{fig4}, the pre-processing stage prepares for the subsequent meta-training and meta-testing. %Since the trace collection is elaborated in \S\ref{finding}, 
 	We focus on the design of bandwidth estimation and filtering mechanism, and the definition of ``network state'' in this subsection. 
 	
 	\textbf{Bandwidth estimation and filtering.}
 	The definition of ``task'' for meta-learning tends to be objective, preventing arbitrary ``task'' changes due to the subjective bitrate selection and policy updates during the training of a given task. The most objective network metrics are available bandwidth and propagation delay, which however cannot be directly measured. Thus, we propose a bandwidth estimation and filtering mechanism based on measurable network metrics. Such mechanism may not be very accurate, but can quickly estimate the trend of ``network state'' changes and use the defined range to mitigate the impact of bias.

 	\begin{figure*}[t]
	\begin{minipage}{0.23\linewidth}
		\centering
		\includegraphics[width=0.96\linewidth]{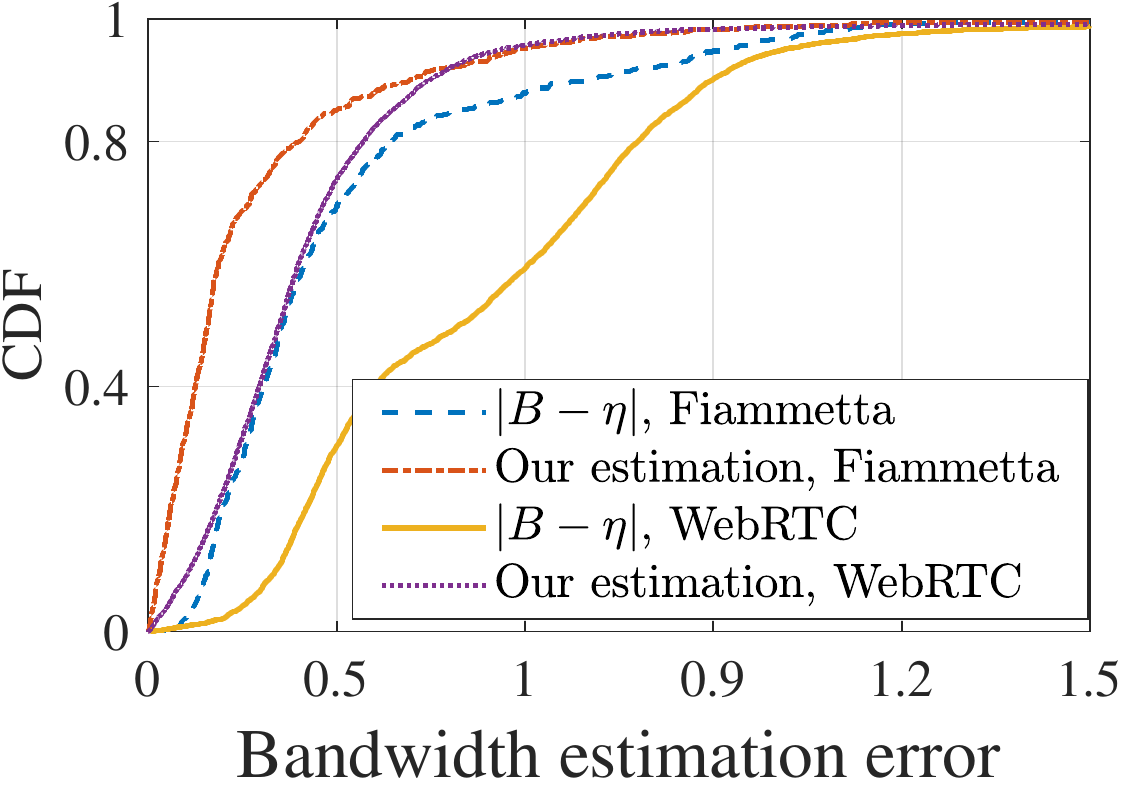}
		\caption{Bandwidth estimation.}
		\label{estimation-value}
	\end{minipage}\enspace
	\begin{minipage}{0.25\linewidth}
		\centering
		\includegraphics[width=0.96\linewidth]{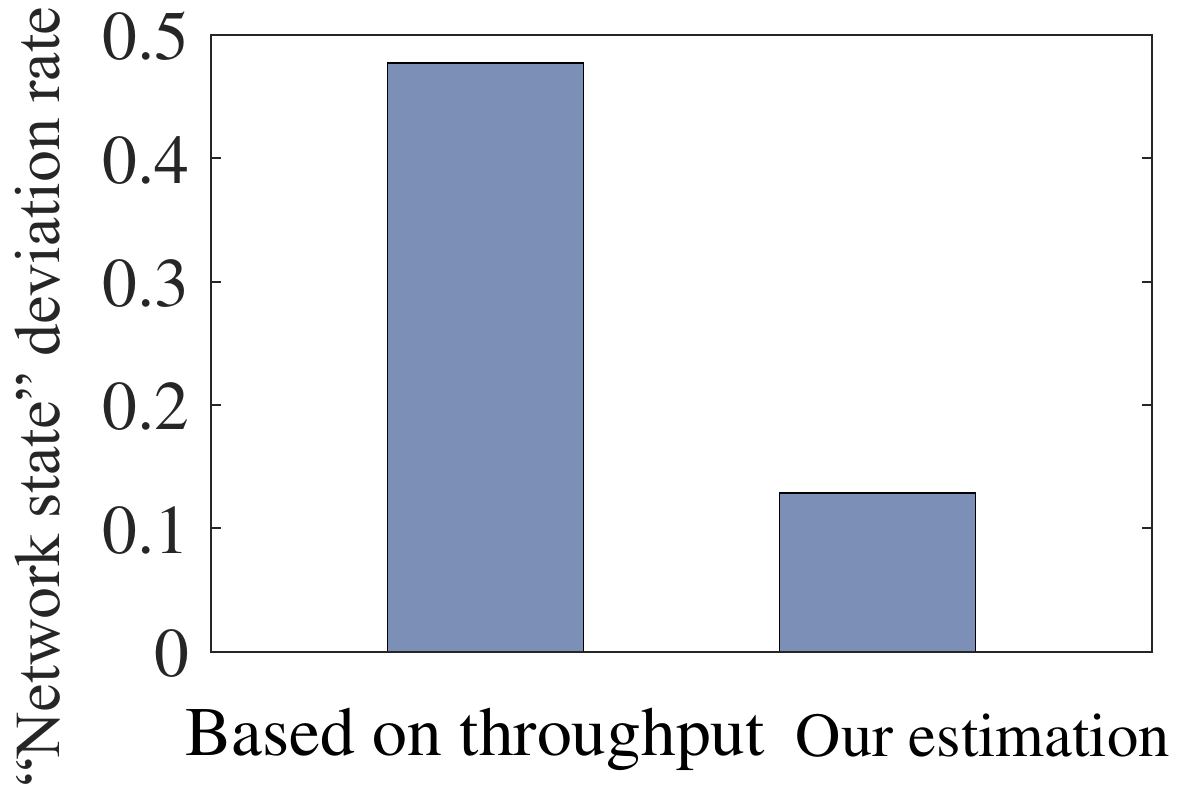}
		\figcaption{``Network state'' estimation.}\label{estimtation-state}
	\end{minipage}\enspace\enspace
	\begin{minipage}{0.49\linewidth}
		\centering
		\includegraphics[width=0.94\linewidth]{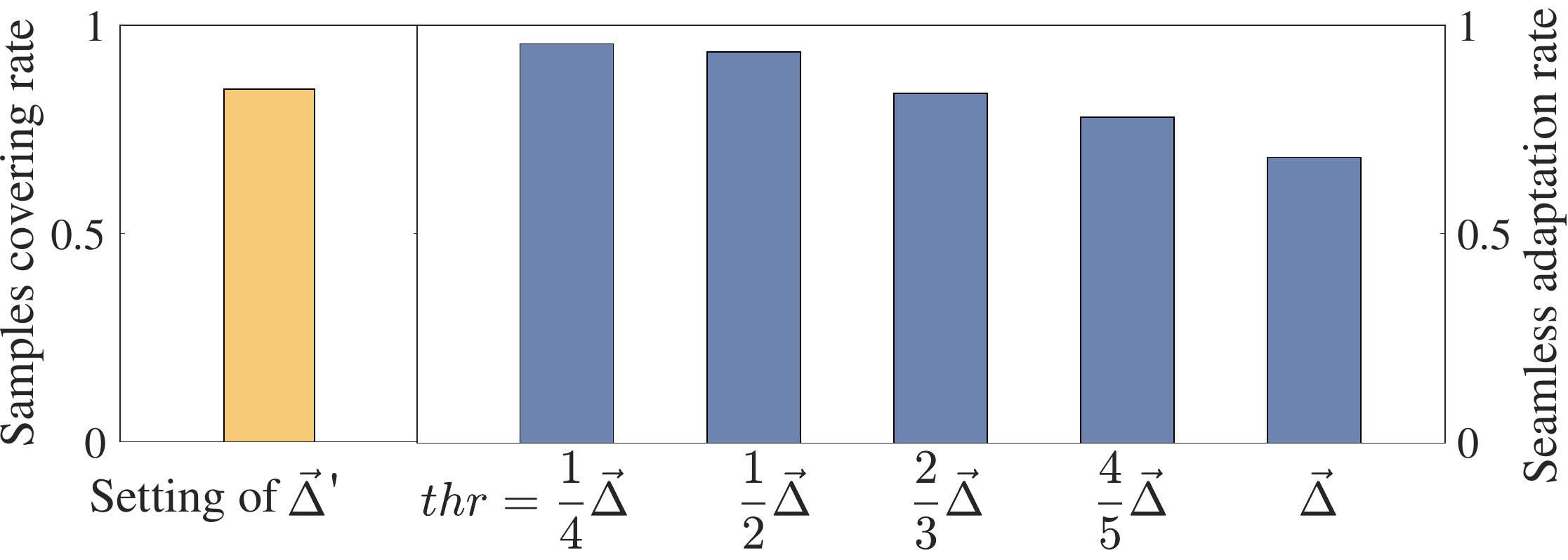}
		\figcaption{Impact of $\boldsymbol{\Delta}'$ and meta-testing $thr$.}\label{seamless-adaptation}
	\end{minipage}\enspace\vspace{-0.3cm}
\end{figure*}

	A basic principle of the bandwidth estimation is that when there exists packet loss or queuing delay caused by congestion, it is at the stage of full pipe and the throughput $\eta_t$ can be regarded as bandwidth~\cite{cardwell2017bbr}. 
	The full pipe is identified (refer to \cite{cardwell2017bbr}) by the following conditions:
	\begin{gather}
		F_t = (l_t> 5\%) \;||\; (d_t> d_{prop,t}+ \sigma (d_{prop})),\\
		d_{prop,t} = \min(d_{t'},\max(d_{prop}))), \\
		t' \in [\max(t-W_d, 0), t],\nonumber
	\end{gather}
	where $l_t$ denotes the packet loss ratio at time $t$, $d_t$ the packet delay,  $d_{prop,t}$ the propagation delay, and $F_t = 1 $ indicates the full pipe. Therein, the threshold of 5\% (refer to \cite{carlucci2016analysis,dong2018pcc}) takes into account the presence of non-congestion packet loss such as lossy wireless links, port flaps on routers, etc., which are not caused by transport-layer bitrate adaptation. The packet delay $d$ is the sum of queuing delay, propagation delay $d_{prop}$ and other smaller values. As path changes on time scales >>
	$d_{prop}$, the $d_{prop,t}$ can be estimated as a running min over a long time window $W_d$~\cite{cardwell2017bbr}. Once $d_t$ exceeds $d_{prop,t}$, we generally assume that there exists queuing delay, i.e., in a full pipe condition. Here, $d_{prop}$ is statistic collected through all $W_d$ windows on the dataset (\S\ref{continuity}), and $\sigma (d_{prop})$ is used to tolerate some irregular delay jitters that are not caused by congestion. Furthermore, the  $\max(d_{prop})$ is configured to avoid $d_{prop,t}$ overestimation at the beginning of video sessions. Then, we propose to estimate bandwidth $\hat{B}_t (Mbps)$ as follows 
	\begin{equation}\label{estimate1}
		\hat{B}_t = \begin{cases}
			\eta_t, & F_t = 1,\\
			\max(\eta_t + pb^2_t, pb^1_t), & F_t = 0.\\
		\end{cases}\\
	\end{equation}
	In unfilled conditions, we estimate by a probing mechanism involving both additive and multiplicative increase, based on $\eta_t$. Here, $pb^2_t$ is set to $\Delta_\mu$, and $pb^1_t$ is calculated by
	\begin{equation}\label{estimate2}
		pb^1_t = \begin{cases}
			\eta_t \times (e^{-\eta_t-1.3} +1), & F_{t-1} = 1,F_t = 0, \\
			pb^1_{t-1} \times (e^{-pb^1_{t-1}-1.3}+1), & F_{t-1} = F_t = 0.
		\end{cases}
	\end{equation}
	%a factor of 0.25 as $\hat{B}_t$. If this condition continues, $\hat{B}_t$ is estimated through further increasing by a factor of 0.25 over $\hat{B}_{t-1}$ or $\eta_t$. 
	Here, $pb^2_t$ is a constant probing value that does not increase with time to  compensate for general phenomenon of $\eta < B$. 
	In contrast, $pb^1_t$ probes with continual multiplicative increase to fit the bandwidth increment. Besides, when this bandwidth estimation and filtering algorithm is used offline, e.g., in a dataset, we can obtain in advance the later $\hat{B}_{later}$ in full pipe and exploit it to adjust $\hat{B}_t$ in unfilled conditions, which is
	\begin{gather}\label{estimate3}
		t' =   \mathop{\arg\max} pb^1_{t'}, \; pb^1_{t'} < \hat{B}_{later}.\\
		\hat{B}_t = \begin{cases}
			\max(pb^1_{t'}-\eta_{t'}, pb^2_t) + \eta_t  \, & t> t'\\
			\hat{B}_t, & else.\\
		\end{cases}
	\end{gather}
	Fig.~\ref{estimation-value} and Fig.~\ref{estimtation-state} demonstrate the effectiveness of our estimation, whether the bitrate is controlled by Fiammetta or GCC~\cite{carlucci2016analysis}. Exploiting our measurement testbed (detailed in \S\ref{sec-implementation}), we can control the bandwidth trace. Our estimation algorithm is able to significantly reduce the estimation error, compared to the original throughput-to-bandwidth gap. In addition, the impact of these errors can be further mitigated by the range $\boldsymbol{\Delta}$, where the deviation rate of ``network state'' estimation is reduced from 0.5 to approximate 0.1.

	%the estimated error amount of this algorithm and compares it with the original throughput and bandwidth difference, and shows it in both cases of running GCC and running Fiammetta

 	\textbf{``Network state'' definition.}  We define the ``task'' as the adaptation to the ``network state'', both denoted as $\Gamma_i, i\in \mathbb{N}^+$. Each $\Gamma_i$ represents network sequences with similar characteristics, i.e., within a specific property cluster of $\{\mu_i$, $\sigma_i$, $\omega_i, d_{prop,i}, \boldsymbol{\Delta}_i \}$, where $\mu_i$, $\sigma_i$, $\omega_i, d_{prop,i}$ are the center and $\boldsymbol{\Delta}_i$ the ranges. All these properties are evaluated within the window $W_r$ (set as 8~s), with 1~s as the unit time. Wherein, the design of $\boldsymbol{\Delta}$ is critical, as the smaller the $\boldsymbol{\Delta}$, the larger improvement in task-specific updates, but the more likely to be affected by meta-testing lags.
 	We therefore set minimum thresholds $\boldsymbol{\Delta}' = \{\Delta'_\mu = 0.2, \Delta'_\sigma = 0.2, \Delta'_\omega =2\}$~(Mbps) according to \S\ref{continuity} and set $\Delta'_{d_{prop}} = 3$~ms in the same way. These $\boldsymbol{\Delta}'$, on the one hand, cover 90\% samples ($\Delta t = 4$~s), as shown in Fig.~\ref{seamless-adaptation}, to ensure seamlAess adaptation during meta-testing, and on the other hand, simplify the complexity of $\Gamma$ space to reduce the difficulty of meta-learning.

%keep 90\% of the time unaffected by lags (i.e., ensuring seamless NN model switching), and on the other hand simplify the complexity of $\Gamma$ space. The thresholds are configured according to the dataset in \S\ref{continuity} and Fig.~\ref{seamless-adaptation} validates the robustness.

 	%这里或下一段还是要说一下，我们保留了大于阈值的range部分，来说适应较大的网络状态变化，使其还是呈现一种动态变化的特征。
 	
 	%4s的事情元测试的时候
 	
 	%还有具体怎么design的元训练的时候解释
 	
	\begin{figure}[t]
	\centering
	\includegraphics[width=0.9\linewidth]{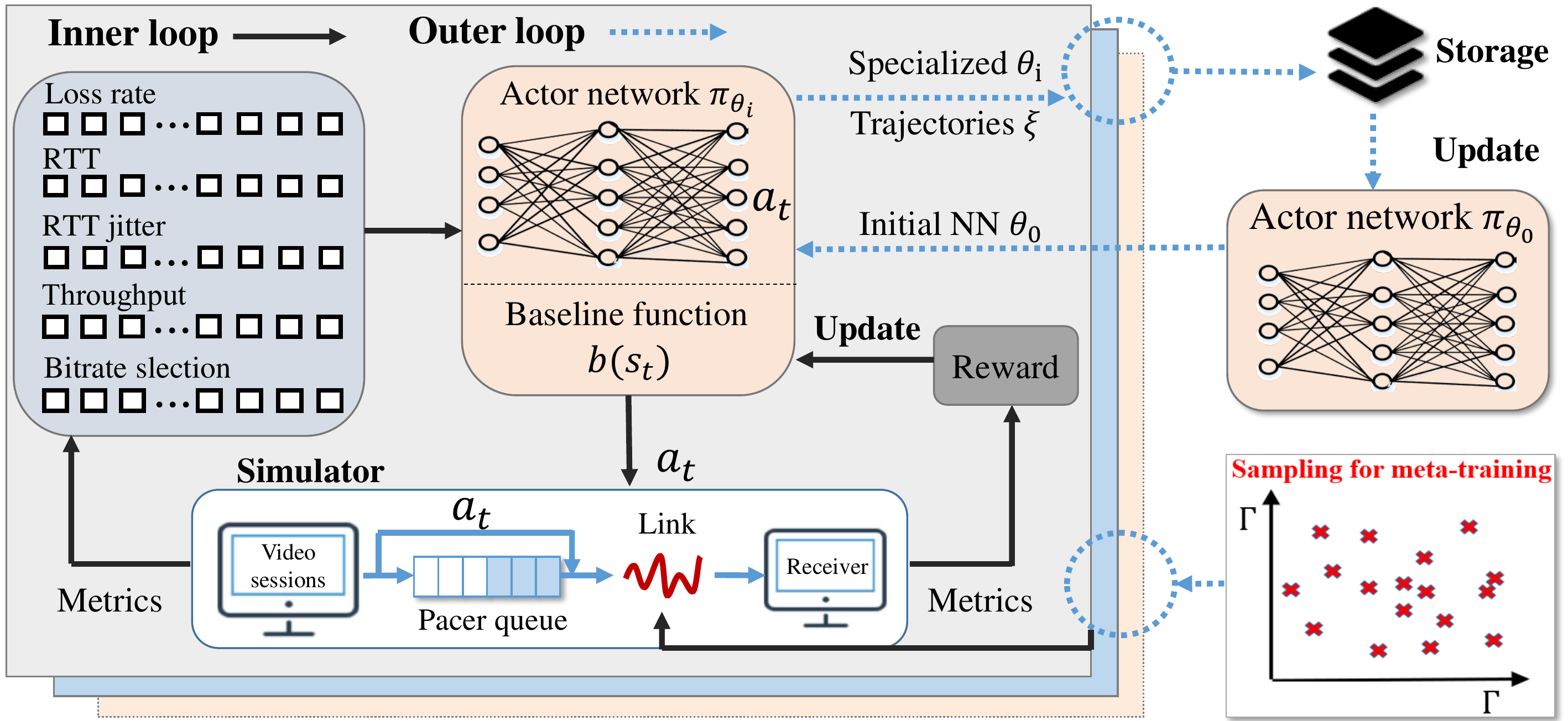}
	\caption{Fiammetta meta-training architecture} \vspace{-0.3cm}
	\label{meta-training-architecture}
	\end{figure}\enspace
 	
 	\subsection{Offline Meta-Training}\label{meta-train}
 	
 	%注意 semi-offline这个说法
 	%波动差好像还没定义
	The meta-training is implemented offline. The key steps and the architecture are depicted in Fig.~\ref{fig4} and Fig.~\ref{meta-training-architecture}, respectively.
 	
 	\textbf{``Network state'' distribution.} Among a series of meta-learning algorithms, we choose model-agnostic meta-learning
 	(MAML)~\cite{finn2017model}. %where the distribution of ``tasks'' is crucial. This is because MAML 
 	MAML essentially trains \textit{initial NN parameters} with high sensitivity on a given ``task'' distribution, allowing for extremely efficient adaptation to  ``tasks'' in few gradient steps. For this reason, we need to obtain the ``network state'' distribution. Specifically, we first apply the bandwidth estimation algorithm (\S\ref{preprocessing}) to the dataset collected in \S\ref{continuity}. Then, we utilize the sliding window $W_r$ to collect $ \{\mu_t ,\sigma_t , \omega_t , d_{prop,t},\boldsymbol{\Delta}_t\}$ of all network sequences, where $\boldsymbol{\Delta}_t = \{\left|\mu_{t}-\mu_{t-\Delta t}\right|,\cdots,\left|d_{prop,t}-d_{prop,t-\Delta t}\right|\}$. %, i.e., the difference of metrics between windows with time interval of $\Delta t$. 
 	We treat each network sequence as the ``network state'' of $\Gamma_i : \{\mu_i = \mu_t, \sigma_i =\sigma_t,\cdots,  \boldsymbol{\Delta}_i =\max(\boldsymbol{\Delta}_t, \boldsymbol{\Delta}')\}$, without the need for additional categories. It is noteworthy that $\Delta t = 4$~s is consistent with that in \S\ref{preprocessing} and \S\ref{continuity} to maintain the short-continuity of ``network states''. Finally, based on the multidimensional ``network states'', we calculate their joint probability densities on the basis of fine-grained partitioning, and then combine interpolation to obtain the probability density function $p(\Gamma) = p(\mu ,\sigma , \omega , \boldsymbol{\Delta})\times p(d_{prop})$.

 	%考虑把metric换成attribute

	\textbf{``Network state'' sampling.} Based on $p(\Gamma)$, we sample ``network state'' $\Gamma_i, i\in \mathbb{N}^+$ to provide tasks for meta-training process, depicted in Fig.~\ref{meta-training-architecture}. Theoretically, the more tasks sampled during meta-training process, the smaller the difference between the distribution of the sampled tasks and $p(\Gamma)$, and the more robust the performance is over $p(\Gamma)$.

	%group probability density function, we sample different categories of network state groups randomly for the tasks required for the meta-training process, according to the number of tasks to be sampled at each outer loop update during meta-training. For example, the sampled network state group categories are

 	\textbf{Generation of network trajectories.} 
 	For each task, sufficient network trajectories are necessary as training samples. However, some ``network states'' $\Gamma$ with small probabilities often encounter the problem of insufficient collected network trajectories during meta-training. Besides, there is also not enough time to obtain network trajectories of new ``network states'' during meta-testing. %g, since our goal is to achieve fast adaptation.
 	Therefore, we explore a second approach which involved generating synthetic network trajectories, in addition to collecting real trajectories.
 	%trace for each 𝑠 with 𝑠’s mean and standard deviation,
 	
 	Given a $\Gamma_i$, we first sample $\mu$, $\sigma$, $d_{prop}$ in their respective %$\left[\max(\mu_i-\Delta_{\mu_i},0),\mu_i+\Delta_{\mu_i}\right]$, and $\sigma$ in $\left[\max(\sigma_i-\Delta_{\sigma_i},0),{\sigma_i}+\Delta_{\sigma_i}\right]$, 
 	ranges, following Gaussian distribution to guarantee that the sampling probability at the center is slightly larger. 
 	Similarly, we sample $d_{prop}$ in the same way. The purpose of this is to provide optimization preference for each $\Gamma_i$. Then, we generate bandwidth samples through two distributions. One is the Gaussian distribution, commonly used in studies~\cite{mao2017neural,akhtar2018oboe} for synthetic trajectory generation, and the other is the beta distribution that is built on a range $\left[0,max \right]$ and can easily achieve asymmetric sampling. Here, $max$ is the maximum $\hat{B}$ in the dataset. 
 	For any given $\mu$ and $\sigma$, if $\left[\mu-3\sigma,\mu+3\sigma \right] \subseteq \left[0,max \right] $ (3$\sigma$ rule~\cite{rule}), we assume a Gaussian distribution for bandwidth samples, otherwise, the beta distribution is adopted as an alternative. 
 	Finally, we repeat the first two steps to obtain enough bandwidth samples, and then arrange them into trajectories in different orders, among which the trajectories that do not satisfy $\Delta_{\mu_i},\Delta_{\sigma_i},\Delta_{\omega_i}$ are filtered out.

 	%Given the network state group category, we randomly sample the mean and standard deviation in the interval, and use the Beta distribution to generate a random value (i.e., the length of the network trajectory at the time of meta-training) that satisfies the mean interval and standard deviation interval and ranges between [0, MAX], and rearrange the sampled values so that the trajectory connected by these random values meets the fluctuation difference interval, the standard deviation interval, and the fluctuation difference interval in each sliding window time. If not, the random values are regenerated until they meet the interval criteria. The above process is repeated until K bandwidth trajectories that meet the interval criteria are generated, forming the class network state group that is needed for the learning task during meta-training

 	\textbf{Meta-training for initial NN model.}  Fig.~\ref{meta-training-architecture} depicts the Fiammetta meta-training architecture. We proceed to describe the customized designs and key algorithms involved.
 	
 	\textit{(\romannumeral1) State and Action.} At any time $t$, the RL agent takes the state %$s_t \in \mathcal{S}$ (i.e., the state space) 
 	$s_t$ as input, the neural network as a function $\pi_{\theta_t}$, and outputs a target bitrate % $a_t \in \mathcal{A}$ (i.e., the state space) 
 	$a_t$ to interact with the interactive video system. The end goal is to find the optimal bitrate adaptation policy, i.e., %$\pi^*_{\theta}:\mathcal{S}\rightarrow \mathcal{A} $ 
 	$\pi^*_{\theta}: s_t \rightarrow a_t$ to maximize QoE. 
 	Specifically, $s_t$ is denoted as $\{\boldsymbol{\eta}_t, \boldsymbol{b}_{t-\Delta t'}, \boldsymbol{l}_t, \boldsymbol{d}_t, \boldsymbol{j}_t\}$, representing sequences of throughput, target bitrate, packet loss ratio, delay, and delay jitter over past 3~s with  $\Delta t' = 0.1$~s as the unite time. 
 	All these metrics can be obtained at the sender via periodic RTCP feedback. After observing $s_t$, the RL agent outputs $a_t$, chosen from 21 discrete actions set $\{-2, -1.8, \cdots, 1.8, 2\}$. Here, $a_t$ represents the scaling factor between two consecutive target bitrate selection, i.e., $b_t =b_{t-\Delta t'} \times e^{a_t} $, like Libra~\cite{du2021unified}. %Upon determining $b_t$, the interaction with environment is achieved by adjusting both codec bitrate and sending rate, as described in Fig.~\ref{fig2}. 

 	\textit{(\romannumeral2) Reward.} Upon determining $a_t$, Fiammetta interacts with the interactive video system by adjusting both codec bitrate and sending rate, and then gets a reward $r_t$ to update $\pi_\theta$. %Since RL seeks to maximize long-term cumulative rewards, 
 	We exploit QoE as the criterion for designing $r_t$, which is
 	\begin{equation}
 		r_t = w_1 \times \eta_t - w_2 \times l_t - w_3 \times d_t - w_4 \times \left| b_t-b_{t-\Delta t'} \right|.
 	\end{equation}
 	Therein, all these metrics are averaged over one unit time, i.e., from $t-\Delta t'$ to $t$. $\left| b_t-b_{t-\Delta t'} \right|$ enforces the codec bitrate smoothness to prevent large frame delay jitter and quality jitter. Referring to recent studies~\cite{zhang2020onrl,zhang2021loki}, we empirically set these four weights to 50, 50, 200 and 20, respectively.
 	
 	\textit{(\romannumeral3) NN structure.} For a start, state sequences are flattened before being fed into networks. The actor network consists of three fully connected layers with 128, 64 and 32 neurons respectively, followed by an activation function. The baseline function $b^{\pi_{\theta}}(s)$ simply implements a linear fit as an average expected reward of $s$ under $\pi_{\theta}$.
 	
 	%这里还需要补充
 	
 	\textit{(\romannumeral4) Training algorithm.}
 	The entire meta-training consists of two parts: inner loop and outer loop. The inner loop is responsible for task-specific optimization, and the outer loop is to obtain an efficient initial NN model, which enables fast adaptation in the inner loop. 
 	
 	During the inner loop, each $\Gamma_i$ contains an initial state distribution $p_i(s_{t})$ and a transition distribution $p_i(s_{t+ \Delta t'}|s_{t}, a_t)$. $\Gamma_i$ is therefore a Markov decision process (MDP). % with a trajectory duration of $N$.  %(default to 16~s).
 	During each gradient update, the RL agent is allowed to query $K$ trajectories $\xi_k = \{s_{\Delta t'}, a_{\Delta t'}, r_{\Delta t'},\cdots, s_{N}, a_{N}, r_{N}\} \;  (k = 1,\cdots, K)$, which are sampled through rollouts by $\pi_{\theta_{i}}$ (initially set as $\theta_0$) on network trajectories (detailed above) and simulators (\S\ref{sec-implementation}).
 	Then, the cumulative reward $R^{\pi_{\theta_{i}}}(s_t, a_t) = \sum_{t' =\frac{t}{\Delta t'}}^{\frac{t+3}{\Delta t'}} \gamma^{(t'-\frac{t}{\Delta t'})}r_{t' \times \Delta t' }$ is used to update $\theta_i$ as follows
 	\begin{align}
 		\mathcal{L}_{\Gamma_{i}}(\theta_i) &=  \mathbb E_{ \xi_k \sim (\pi_{\theta_i}, p_i)} \left[\sum_{t}^{\xi_k}  \hat{A}^{\pi_{\theta_{i}}}(s_t, a_t)\right], \label{loss}\\
 		\theta_{i} &\leftarrow \theta_{i} + \alpha \Delta_{\theta_{i}} \mathcal{L}_{\Gamma_{i}}(\theta_i), \label{update}
 	\end{align}
 	%\begin{gather}
 	%	\mathcal{L}_{\Gamma_{i}}(\theta_i) =  \mathbb E_{ \xi_k \sim (\pi_{\theta_i}, p_i)} [\sum_{t' = 1}^{(N-3)/\Delta t' }  \hat{A}^{\pi_{\theta_{i}}}(s_{t'\times \Delta_t}, a_{t'\times \Delta_t})], \label{loss}\\
 	%	\theta_{i} \leftarrow \theta_{i} + \alpha \Delta_{\theta_{i}} \mathcal{L}_{\Gamma_{i}}(\theta_i), \label{update}
 	%\end{gather}
	where $\mathcal{L}$ is the loss function, $\alpha$ the learning rate, $b^{\pi_{\theta_{i}}} (s_t) $ the average expected reward at $s_t$ under $\pi_{\theta_{i}}$, $\hat{A}^{\pi_{\theta_{i}}} (s_t, a_t) = R^{\pi_{\theta_{i}}} (s_t, a_t) - b^{\pi_{\theta_{i}}} (s_t) $ the advantage function, representing the extra benefit from a certain $a_t$. % compared to the expected reward across various actions. 
	Besides, ``3~s'' is the empirically configured time that Fiammetta considers for the future. 
 	When adapting to $\Gamma_i$, the NN parameters evolve from $\theta_0$ to $\theta_i$ according to Eq.~\eqref{update} through only 3 gradient updates that are sufficient for convergence. 
 	
 	The outer loop further improves the performance of $\pi_{\theta_i}$ by updating initial $\theta_0$ following PPO algorithm~\cite{schulman2017proximal}. During each round of outer loop, Fiammetta samples $M$ tasks from $p(\Gamma)$, updates to corresponding $\{\theta_i\}_{i = 1}^M$ in the inner loop, and evaluates by resampling $K$ trajectories $\{\xi_{k,i}\}_{k = 1}^K$ on each task using $\pi_{\theta_i}$. %After recalculating $\mathcal{L}_{\Gamma_{i}}(\theta_i)$ based on $\pi_{\theta_i}$, 
 	Then, Fiammetta obtains the loss function $\mathcal{L}$ and updates $\theta_0$ by
 	\begin{align}
		\mathcal{L}^{\theta'_{0}} (\theta_{0}) =  & \mathbb E^{i \sim p(\Gamma)}_{ \xi_{k,i} \sim (\pi_{\theta_i}, p_i)} \bigg[ \sum_{t}^{\xi_{k,i}} \min \left(  \delta ^{\theta'_{0}}(\theta_{0}) \hat{A}^{\pi_{\theta_{i}}}(s_t, a_t), \right. \notag\\   
		&\left. clip (\delta^{\theta'_{0}}  (\theta_0),1-\epsilon, 1+\epsilon) \hat{A}^{\pi_{\theta_{i}}}(s_t, a_t)  \right) \bigg], \label{outer-loop1} \\ 
		\theta_{0} \leftarrow &\theta_{0} + \beta \Delta_{\theta_{0}} \mathcal{L}^{\theta'_{0}} (\theta_{0}), \label{outer-loop2}
	\end{align}
	where $\theta'_0$ is the old initial parameters before each round of outer-loop update, and $\delta^{\theta'_{0}}(\theta_{0}) = \frac{\pi_{\theta_{0}}(s_t, a_t)}{\pi_{\theta_{old, 0}}(s_t, a_t)} $ represents the ratio of a new policy and its old one. The basic idea is to make the update of $\theta_0$ smoother and avoid gradient oscillations by clipping $\delta^{\theta'_{0}}(\theta_{0})$ values that are out of $[1-\epsilon, 1+\epsilon]$.

	 \begin{figure}[t]
		\centering
		\includegraphics[width=0.97\linewidth]{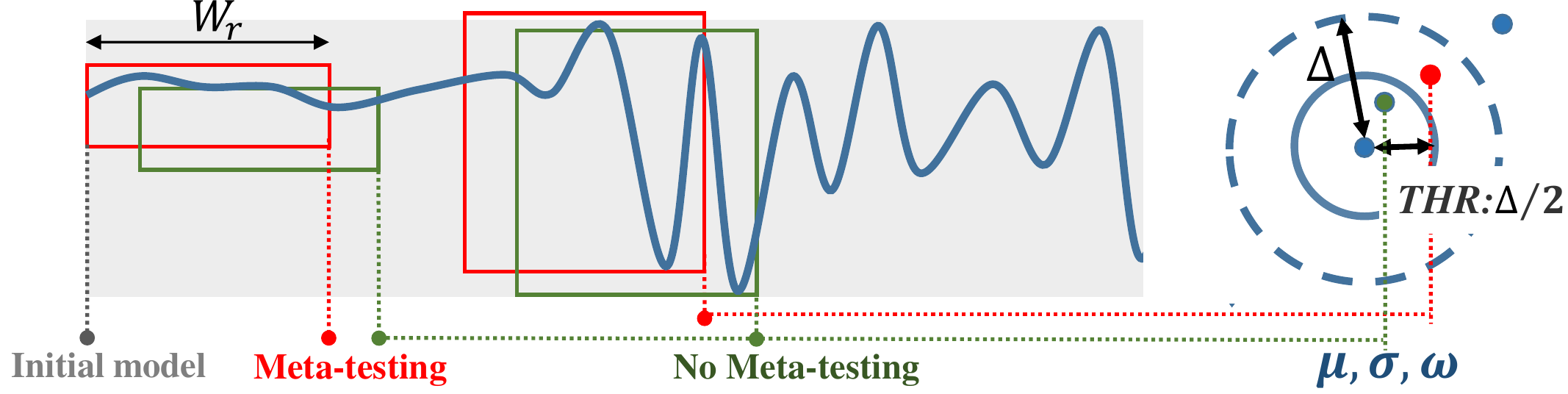}
		\caption{Fiammetta meta-testing design.}\vspace{-0.3cm}
		\label{fig:meta-test}
	\end{figure}

 	\subsection{Online Meta-Testing}\label{meta-testing}
 	
	Based on the initial NN model, Fiammetta achieves seamless adaptation to time-varying ``network states'' through online meta-testing, the steps of which are shown in Fig.~\ref{fig4} and Fig.~\ref{fig:meta-test}.
 	
 	\textbf{``Network state'' monitoring.} Fiammetta monitors the real-time ``network state'' by computing $\mu_t, \sigma_t, \omega_t$ and $d_{prop,t}$ of the network sequence within a sliding window $W_r$ over past $8$~s. It is noteworthy that the bandwidth $\hat{B}$ and $d_{prop}$ are estimated following the steps in \S\ref{preprocessing}, and the sliding window is moved at a granularity of 1~s for real-time assurance.
 	
 	\textbf{Meta-testing activation.} The target of meta-testing activation is to achieve a seamless adaptation between ``network state'' changes. Therefore, it cannot wait until the detected attributes, such as $\mu_t, \sigma_t, \omega_t, d_{prop,t}$ are out of the ranges $\boldsymbol{\Delta}_{last}$ of last ``task''. As an alternative, we set the activation threshold to $\frac{\boldsymbol{\Delta}_{last}}{2}$ according to Fig.~\ref{seamless-adaptation}, i.e., each of $\left|\mu_t-\mu_{last}\right| > \frac{\Delta_{\mu,last}}{2}$, $\left|\gamma_t-\gamma_{last}\right| > \frac{\Delta_{\gamma,last}}{2}$,  $\cdots$ will activate meta-testing. Then, the seamless adaptation is ensured by alleviating the effect of meta-testing lag,  shown in Fig.~\ref{fig:meta-test}. Note that $\boldsymbol{\Delta}$ is also required to be compliant with this lag. As meta-testing (i.e., inner loop with 3 gradient updates) takes roughly 2~s according to our statistics,  $\boldsymbol{\Delta}$ needs to cover at least 4~s of network variations, which is the reason for configuring $\Delta t = 4$~s (\S\ref{continuity}) and obtaining $\boldsymbol{\Delta}'$ (\S\ref{meta-train}) based on this.
 	
 	%这里需要加上时间窗口的说明

 	\textbf{Meta-testing.} Consistent with the inner loop, meta-testing is responsible for adapting to varying "network states" detected in real time. Upon activation, the new ``network state'' $\Gamma_i: \{\mu_i = \mu_t, \sigma_i =\sigma_t,  \omega_i = \omega_t,  \boldsymbol{\Delta}_i =\max(\boldsymbol{\Delta}_t, \boldsymbol{\Delta}')\}$ is generated, where $\boldsymbol{\Delta}_t  = \{\left|\mu_{t}-\mu_{last} \right|, \left|\sigma_{t}-\sigma_{last} \right|, \left|\omega_{t}-\omega_{last} \right|\}$. Then, we follow the workflow of inner loop by first sampling $K$  trajectories of $\Gamma_i$, again obtained through $K$ rollouts on synthetic network trajectories (\S\ref{meta-train}) and simulators (\S\ref{sec-implementation}). Then, the specialized NN parameters $\theta_i$ for $\Gamma_i$ can be obtained through 3 gradient updates based on Eq.~\eqref{loss}-\eqref{update}.

 	\section{Implementation}\label{sec-implementation}

\begin{figure}[t]
 		\centering
 		\includegraphics[width=0.94\linewidth]{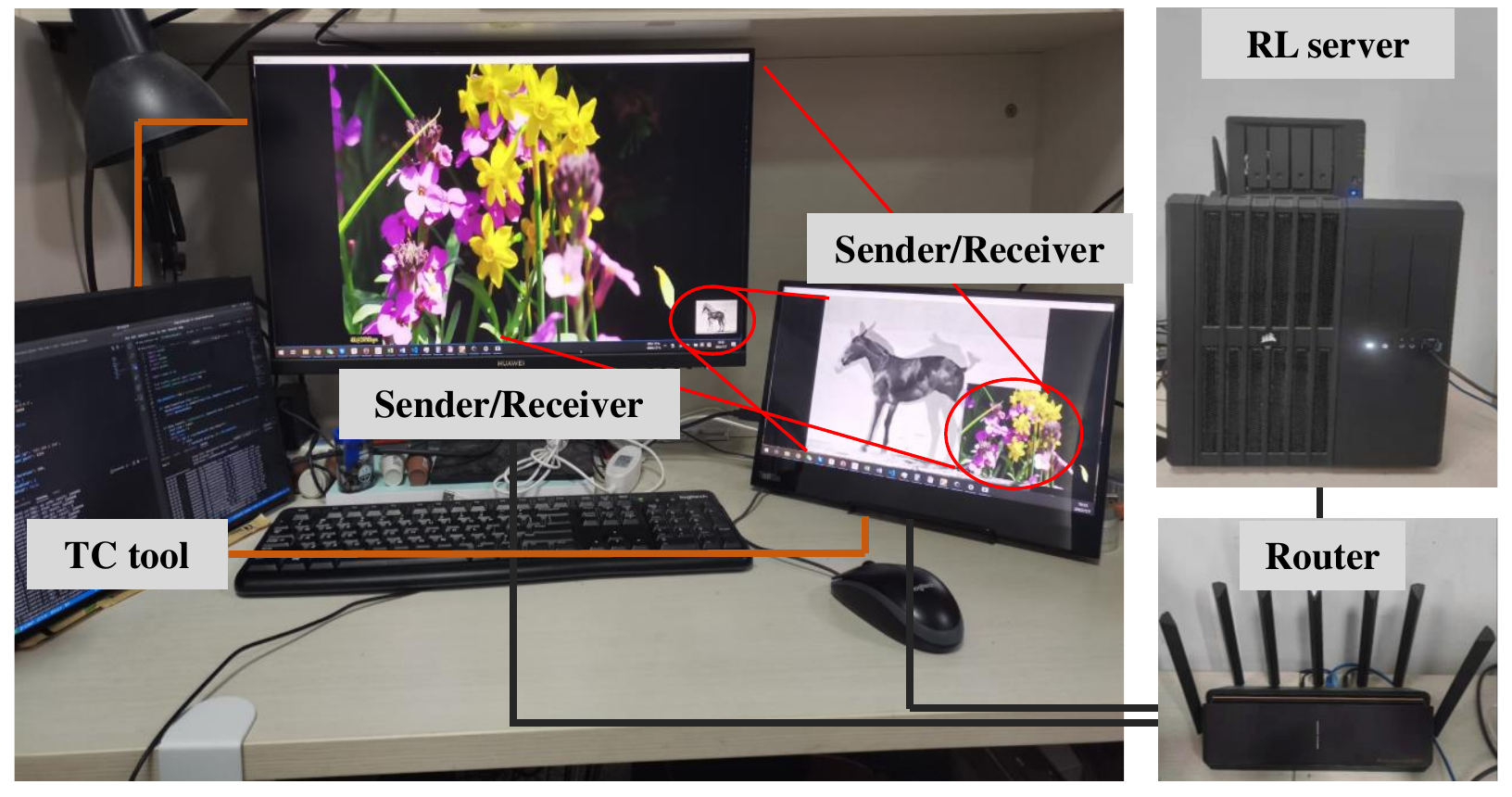}
 		\caption{Testbed setup.}\vspace{-0.3cm}
 		\label{testbed}	
 	\end{figure}

\begin{figure}[t]
 		\centering
 		\includegraphics[width=0.85\linewidth]{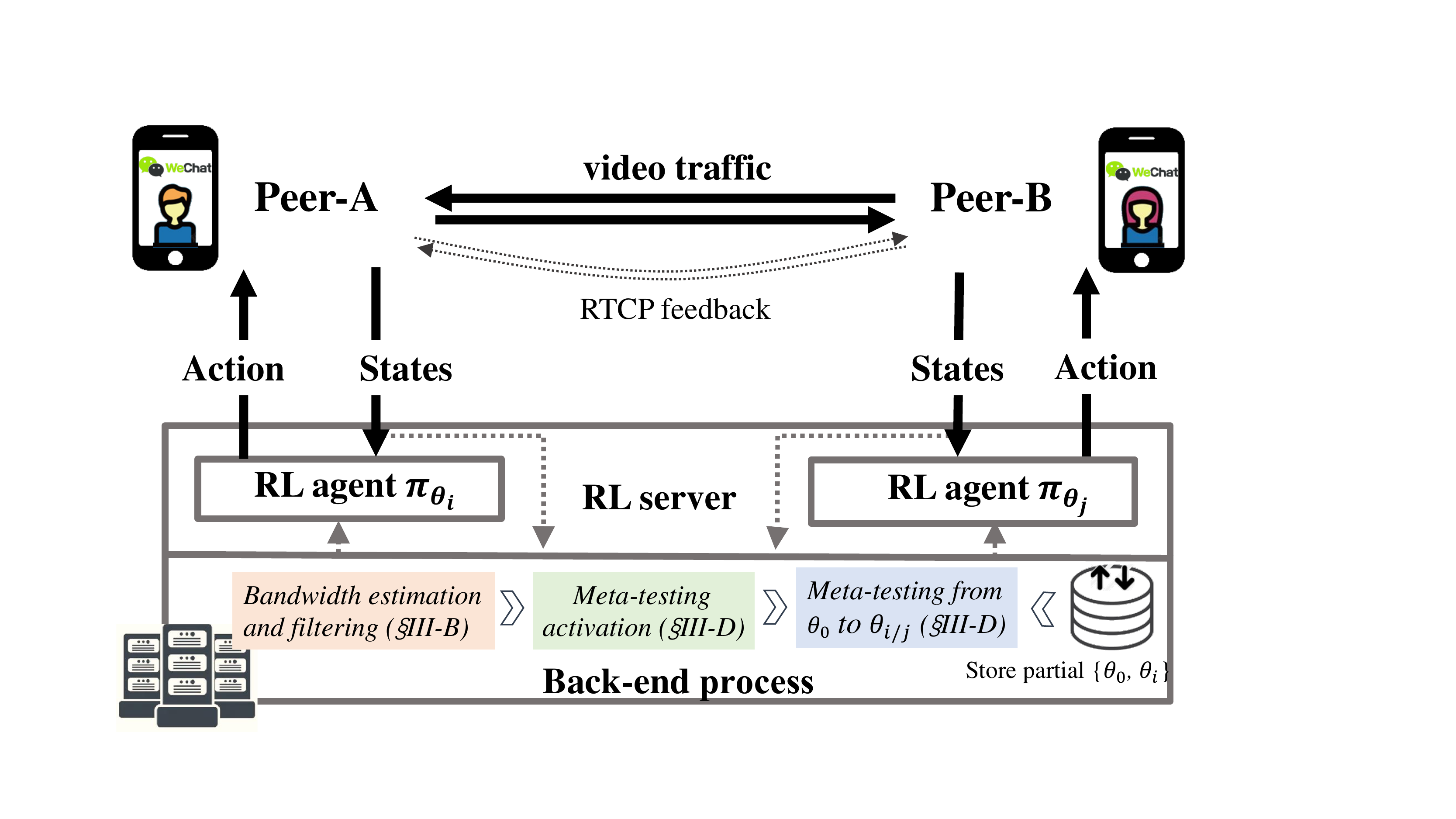}
 		\caption{System implementation of Fiammetta.}\vspace{-0.3cm}
 		\label{fig-implementation}	
 	\end{figure}

 	\textbf{Testbed implementation.} We build an end-to-end measurement testbed and exploit the real-world network traces (detailed in \S\ref{finding}) sponsored by WeChat for Business APP to test the performance of Fiammetta and baseline algorithms. As shown in Fig.~\ref{testbed}, the testbed mainly consists of two PCs running WebRTC as a video traffic transceiver pair and one PC controlling network link through the TC (traffic control) tool~\cite{tc}. Besides, we implement Fiammetta and learning-based baseline algorithms on a remote RL server, and the video transceiver pair is connected to the RL server via an additional router to query the target bitrate. The RL server is a desktop equipped with Intel Core i7-9700K CPU, Geforce RTX 1080Ti GPU, 32-GB memory and Ubuntu 18.04. We implement Fiammetta with PyTorch version 1.10.2. %The model has a small size of XXKB.
 	Fig.~\ref{fig-implementation} further shows the detailed system implementation of Fiammetta, which works by replacing the bitrate control module in interactive video system. 
	At run time, the sender offloads network metrics (obtained from RTCP feedback) to the RL server in real-time and adopt the bitrate decision $a_t$ selected by the RL model $\pi_{\theta_i}$. %In addition, we separately deploy back-end servers for online meta-testing, on which three related modules are implemented. Once the meta-testing is completed, the old $\theta_{old}$ is replaced by the new $\theta_{new}$. 
	Moreover, we detach meta-testing from the front-end query service to ensure responsiveness. Specifically, we set up a back-end process to receive the transition tuples from the front-end and conduct meta-testing without disturbing it. Once meta-testing is completed, the old NN parameters $\theta_{old}$ in the front-end will be replaced by the new parameters $\theta_{new}$. To further reduce computation overhead, we dynamically store partial specialized NN parameters $\theta_i$ on the server according to the access frequency, in addition to initial NN parameters.

 	\textbf{Simulator.} Our simulator consists of two modules: \textit{(\romannumeral1)} A video compression module that compresses real video to different bitrates of $\{0.1, 0.2,\cdots , 2.5\}$~Mbps, and records each frame size to faithfully simulate frame size jitters. At run time, this module continuously outputs frame sizes of the video that has the closest bitrate to the target. \textit{(\romannumeral2)} A transport module that packages frame-sized random data into RTP packets and faithfully simulates the pacer mechanism to send packets into a simulated network path. The bandwidth and propagation delay are configured according to the WeChat for Business traces for Fiammetta training.

 	\section{Evaluation}\label{evaluation}
 	%In this section, we evaluate Fiammetta from three aspects: \textit{(\romannumeral1)} We 
 	%first test Fiammetta's overall performance and demonstrate its superior over state-of-the-art algorithms (\S\ref{overall performance}); \textit{(\romannumeral2)} We validate the strong adaptability of Fiammetta to different ``network states''~(\S\ref{different network states}); \textit{(\romannumeral3)} We compare Fiammetta with its initial model (without online meta-testing) to understand the effectiveness of Fimametta's few-shot learning process (\S\ref{without meta-testing}).
 	\subsection{Methodology}
 	\textbf{Trace-driven testbed experiments.} 	We train and test Fiammetta \& baseline algorithms on the same datast (detailed in \S\ref{finding}) with 75\% the training set and 25\% the test set. As the network bandwidth is hard to accurately measure at fine grain, we use throughput as bandwidth for experiments. This certainly simplifies the experiment by getting a precise ``network states'' distribution $p(\Gamma)$. However, we want to make it clear that our design can handle real-world deployments without explicit bandwidth information. %To guarantee the robustness, we adopt 10 videos with different contents ranging from landscapes to animations for training and testing. All the results in this section are obtained through trace-driven testbed experiments. 
 	
 	\begin{table*}[t]
 		\renewcommand{\arraystretch}{1.25}
 		\centering
 		\caption{Overall performance regarding application and transport-layer metrics ($Mean \pm Std\,Dev$) }
 		\begin{tabular}{m{1.2cm}<{\centering}|m{1.1cm}<{\centering}m{1.1cm}<{\centering}m{1.4cm}<{\centering}m{1.23cm}<{\centering}m{1.25cm}<{\centering}|m{1cm}<{\centering}m{1.2cm}<{\centering}m{1.2cm}<{\centering}m{1.45cm}<{\centering}|m{1.3cm}<{\centering}}
 			\toprule[1.5pt]
 			\multirow{2}{*}{\makecell[c]{~\\ ~\\\textbf{Algorithms}}} & \multicolumn{5}{c|}{\textbf{Application-layer metrics}}&\multicolumn{4}{c|}{\textbf{Transport-layer metrics}} & \multirow{2}{*}{\makecell[c]{~\\ \textbf{QoE/}\\\textbf{Reward}}}\\
 			\cline{2-10}
 			\multicolumn{1}{c|}{} & FPS & \makecell[c]{PSNR \\ (dB)} & \makecell[c]{Stalling rate \\ (FPS <12, \%)} & \makecell[c]{Frame delay\\ (ms)} & \makecell[c]{Frame delay\\ jitter (ms)} & \makecell[c]{RTT\\ (ms)} & \makecell[c]{Loss rate \\ (\%)}  & \makecell[c]{Throughput\\ (Mbps)} & \makecell[c]{Bitrate jitter\\ (Mbps/10min)} & \multicolumn{1}{c}{}\\
 			\midrule[0.85pt]
 			
			GCC~\cite{carlucci2016analysis}  & $29.41 \pm 4.41$   & $33.17 \pm 7.98$  & $0.83\pm5.27$  & $\boldsymbol{160.23} \pm \boldsymbol{45.42}$  & $ \boldsymbol{5.25} \pm \boldsymbol{10.35} $  & $68.93\pm110.27$  &  $1.68 \pm 1.01 $ & $0.74\pm0.23$ & $1.09\pm2.04 $& 22.85\\

			%0.25x-----------------
			OnRL~\cite{zhang2020onrl} & $29.02 \pm 4.29$ & $34.12\pm 7.32$  &  $0.96\pm3.55$ &  $170.68 \pm 36.32$ &  $ 5.53 \pm 10.82 $ &  $73.95\pm119.01$ &  $2.00 \pm 1.33 $& $ 0.83\pm0.25$ & $ \boldsymbol{0.92}\pm \boldsymbol{5.20}$ & 26.05\\

			Loki~\cite{zhang2021loki}  & $29.04\pm4.36 $ & $33.82\pm 6.55$  & $0.80\pm2.32$  &$165.21 \pm 38.34$   & $ 5.343\pm 10.92 $ &$69.29\pm97.86$  &$1.88 \pm 1.25 $  &$0.78\pm0.22$ & $1.20\pm1.81$& 24.74\\

			Fiammetta  & $\boldsymbol{29.51}\pm\boldsymbol{3.41}$  & $\boldsymbol{34.51} \pm \textbf{7.27}$  & $\boldsymbol{0.75}\pm\boldsymbol{2.31} $ & $162.53 \pm 30.24$  & $ 5.27 \pm 9.63 $  & $\boldsymbol{68.51}\pm \boldsymbol{63.86}$  &$\boldsymbol{1.65} \pm \boldsymbol{1.25} $  & $\boldsymbol{0.86}\pm \boldsymbol{0.24}$ & $1.09\pm2.16$& $\boldsymbol{28.94}$\\
 			\bottomrule[1.5pt]
 		\end{tabular}\vspace{-0.2cm}
 		\label{table:overall performance}
 	\end{table*}
 	
 	\textbf{Baseline algorithms} are listed in the following  %We compare Fiammetta with three state-of-the-art algorithms:
 	\begin{itemize}
 		\item [\textit{(\romannumeral1)}] 
 		\textit{GCC}~\cite{carlucci2016analysis}, as an official congestion control algorithm, is widely used in mainstream interactive video systems, such as WeChat, Google Hangouts, etc. The core idea is to adopt a combination of loss-based and delay-gradient-based methods to avoid congestion and adjust bitrates. 
 		\item [\textit{(\romannumeral2)}]
 		\textit{OnRL}~\cite{zhang2020onrl} is the first online-RL-based adaptation algorithm for interactive video systems. We train it with the same training set directly on our testbed, spending about 12 days with 291 hours, and implement online adaptation during the 4-day-long test period of 100 hours.
 		\item [\textit{(\romannumeral3)}]
 		\textit{Loki}~\cite{zhang2021loki} proposes to fuse the rule-based GCC with the RL-based algorithm at the feature level to improve the long-tail performance. Following Loki, we ``blackboxify'' GCC, integrate it with OnRL, and then retrain the new NN on our testbed with the same training set. The online adaptation is also performed during testing process.
 		
 	\end{itemize}
 
 	%\textbf{Evaluation metrics.} Our evaluation cover both application- and transport-layer metrics: \textit{(\romannumeral1)} The application-layer metrics include FPS, PSNR, stalling rate (ratio of FPS < 12), frame delay and delay jitter, where PSNR is indicated by QP that can be measured in real time~\cite{zhang2021loki}. \textit{(\romannumeral2)} The transport-layer metrics include RTT, loss rate, throughput and bitrate jitter, all of which have direct impact on application-layer metrics.

 	\subsection{Comparison with Baseline Algorithms} \label{overall performance}
 	The comparison results are summarized in Table~\ref{table:overall performance}. 
 	
	\textbf{Application-layer metrics.} Fiammetta significantly outperforms all three competing algorithm: \textit{(\romannumeral1)} Compared to GCC, Fiammetta improves PSNR by 1.34~dB %, increases FPS by 4.0\%, 
	and decreases stalling rate by $9.6\%$. More importantly, Fiammetta is also comparable to the conservative GCC in terms of FPS, frame delay and jitters. \textit{(\romannumeral2)} In comparison to learning-based OnRL and Loki, Fiammetta cuts stalling rate by 21.9\% and 6.3\%, respectively, and is also better in frame quality, with a little improvements in both PSNR and FPS. Also, the frame delay and delay jitter of Fiammetta exhibit better performance than Loki, OnRL.

 	\textbf{Transport-layer metrics.} Similarly, Fiammetta achieves remarkable gains in most transport-layer metrics: \textit{(\romannumeral1)} Consistent with application-layer metrics, Fiammetta effectively improves throughput by 3.6\%, 10.3\%, 16.2\% over OnRL, Loki and GCC, respectively. \textit{(\romannumeral2)} Meanwhile, the RTT of Fiammetta significantly drops by 1.1\%, 7.4\%, synchronized with loss rate reduction of 12.3\%, 17.5\%, compared to Loki and OnRL. These metrics are at the same level as the conservative GCC. \textit{(\romannumeral3)} While the bitrate jitter of Fiammetta is slight worse than OnRL, it performs comparable to GCC and better than Loki.

 	\textbf{QoE/Reward.} Table~\ref{table:overall performance} further demonstrates the  QoE/reward to provide an overall evaluation. For fairness comparison, all RL-based algorithms are trained using the same QoE/Reward setting defined in \S\ref{meta-train}.
 	We can notice that Fiammetta outperforms OnRL by 11.1\% in QoE, and the gap gradually increases to 17.0\% and 26.7\% when compared to Loki and GCC, respectively. These results validate the overall superiority of Fiammetta, by achieving a better dynamic balance between conflicting metrics such as delay, packet loss, smoothness and throughput to maximize interactive video QoE.
 	
 	\begin{figure}[t]
 		\centering
 		\includegraphics[width=1\linewidth]{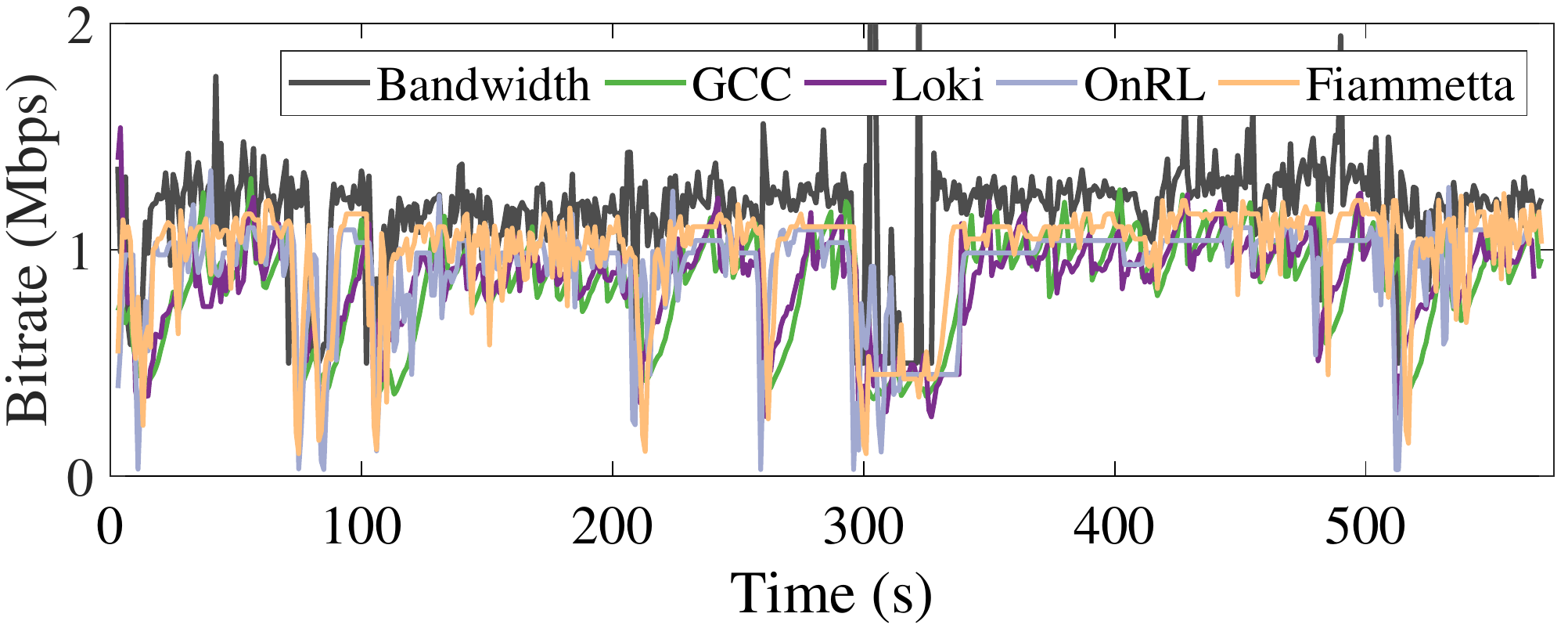}
 		\caption{A microscopic showcase of a 600-second session.}\vspace{-0.4cm}
 		\label{microscopic}	
 	\end{figure}
 
 	\textbf{A microscopic showcase.} We find that Fiammetta's gains stem from its ability to quickly generate and switch to strategies that fit the current ``network state''. Fig.~\ref{microscopic} exhibits a representative 600-second session. \textit{(\romannumeral1)} When the bandwidth is relatively stable, Fiammetta follows closely, maximizing bandwidth utilization with minimum magnitude of probing. The reason is that Fiammetta can always have rough priori assumptions about ``network states''. At run time, when the ``network state'' is roughly detected, Fiammetta can quickly generate strategies that fit the range of ``network state'', eliminating many substantial trial-and-error behaviors.
 	In comparison, GCC suffers from periodic bitrate degradation due to its AIMD-based probing mechanism, while OnRL exhibits frequent overshoots. Affected by the integration with GCC, Loki also encounters unnecessary bitrate drops, when GCC bitrate goes down and occupies much larger impacting factor, resulting in a loss of bandwidth utilization. \textit{(\romannumeral2)} When the bandwidth fluctuates dramatically (280-370~s), Fiammetta can also benefit from the priori assumptions and generate relatively conservative strategies compared to OnRL, yielding an RTT comparable to GCC and with higher throughput. 

 	 \begin{figure*}[t]
		\centering
		\includegraphics[width=0.99\linewidth]{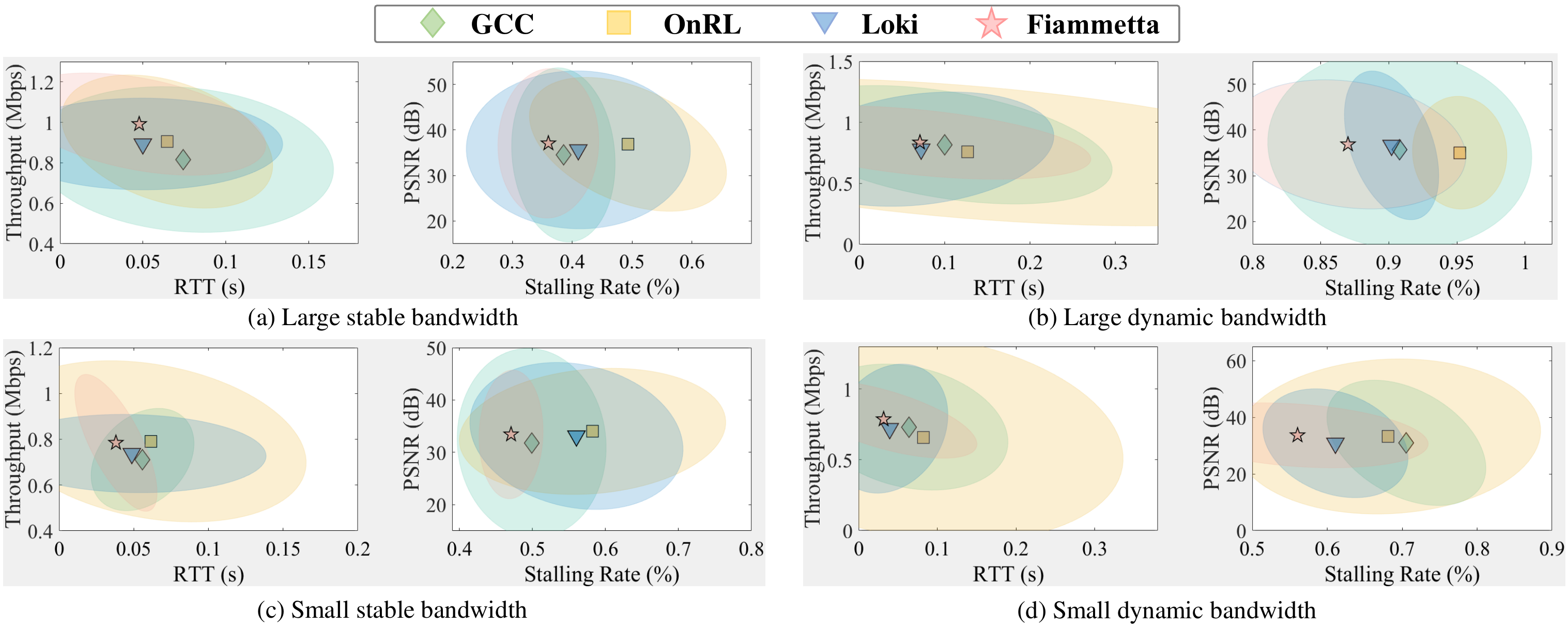}
		\caption{Fiammetta's adaptability to different ``network states'', compared to baseline algorithms.}\vspace{-0.4cm}
		\label{qipao}	
	\end{figure*}
 	
 	%Concerto outperforms all three competing algorithms, with  

 	\subsection{Comparison across Different ``Network States''} \label{different network states}
 	%We further conduct a breakdown analysis to evaluate Fiammetta across large/small stable/dynamic bandwidth, with results depicted in Fig.~\ref{qipao}.
 	We further evaluate Fiammetta across large/small stable/dynamic bandwidth, with results depicted in Fig.~\ref{qipao}.
 	%In this evaluation, all results across different ``network states'' are directly extracted from \S\ref{overall performance}.
 	
 	\textbf{Large stable bandwidth.} We first evaluate Fiammetta in the highest-quality ``network state'' with large stable bandwidth ($\mu \in [1,3] $ and $\sigma \in [0,0.1]$). %Depicted in Fig.~\ref{higha} and Fig.~\ref{highb}, 
 	%Fiammetta %outperforms all competing algorithms, with 
 	%achieves remarkable throughput gains by xx\%, xx\%, xx\%, and PSNR gains by xx~dB, xx~dB, xx~dB, compared to Loki, OnRL and GCC. 
 	Fiammetta achieves noticeable video clarity improvements, boosting throughput by 9.5\%-21.4\%, and PSNR by 0.2~dB-2.6~dB across three baseline algorithms. 
 	Meanwhile, Fiammetta also achieves better performance in RTT and stalling rates.
 	This results further validate that priori rough assumptions about ``network states'' help Fiammetta to achieve better performance.
 	%In terms of RTT and stalling rate, Fiammetta achieves nearly the same performance as GCC, and with reductions in RTT of 3\%, 5\% and in stalling rate of xx\%, xx\% over Loki and OnRL. 

 	%We analyze that the gains come from prior assumptions about ``network states''. Once the ``network state'' is roughly detected, Fiammetta can quickly generate policies that fit the ``network state'', eliminating many substantial trial-and-error behaviors.
 	
 	\textbf{Large dynamic bandwidth.} For the ``network state'' with large but dynamic bandwidth ($\mu \in [1,3] $ and $\sigma \in [0.3,0.5]$), Fiammetta  %outperforms all competing algorithms, with 
 	still achieves throughput gains by 2.2\%-10.3\%, and PSNR gains by 0.2~dB-1.9~dB. %, across three baseline algorithms. 	
 	%Fiammetta achieves the greatest video clarity improvements, boosting throughput by xx\%, xx\% and xx\%, and PSNR by xx~db, xx~db and xx~db than Loki, OnRL and GCC. 
 	%While the RTT is slightly larger than GCC, the impact on FPS is very slight, let alone leading to more stalling rates. 
 	Meanwhile, both RTT and stalling rate exhibit noticeable reductions compared to baseline algorithms 
 	%Meanwhile, Fiammetta achieves better performance in RTT and FPS compared to all three baseline algorithms. 
 	This shows that Fiammetta can quickly switch to a relatively conservative strategy when bandwidth fluctuates, to avoid much latency increase.
 	
 	\textbf{Small stable bandwidth.} When adapting to ``network state'' with small stable bandwidth ($\mu \in [0.5,1] $ and $\sigma \in [0,0.1]$), 
 	Fiammetta achieves stalling rate comparable to GCC and significantly better than Loki and OnRL. Moreover, Fiammetta's PSNR and throughput are in between OnRL and Loki, achieving global optimum in QoE.
 	%Fiammetta achieves 6.4\%-10.3\% throughput gains, and hence 3.0~dB-1.7~dB PSNR gains over most baseline algorithms. Meanwhile, it decreases RTT by 18.3\%-37.2\% and stalling rate by 5.7\%-19.2\%.
 	%Meanwhile, the RTT is much less than OnRL and Loki (xx\% and xx\%).
 	Such results suggest Fiammetta has essentially developed strategies for small stable bandwidth, instead of blindly probing, causing frequent overshoots.
 	
 	\textbf{Small dynamic bandwidth.} The small dynamic bandwidth ($\mu \in [0.5,1] $ and $\sigma \in [0.3,0.5]$) is considered by us as the worst ``network state'', but Fiammetta still achieves promising results in this case. Instead of blindly seeking to fit the bandwidth fluctuation curve, Fiammetta adopts a relatively conservative strategy. It achieves 7.5\%-19.8\% throughput gains, and hence 0.4~dB-2.8~dB PSNR gains, with reductions in stalling rate of 8.3\%-20.6\% over three baseline algorithms. 
 	
 	\subsection{Comparison of Fiammetta w/ and w/o Meta-Testing} \label{without meta-testing}
\begin{figure}[t]
		\centering
		\includegraphics[width=0.98\linewidth]{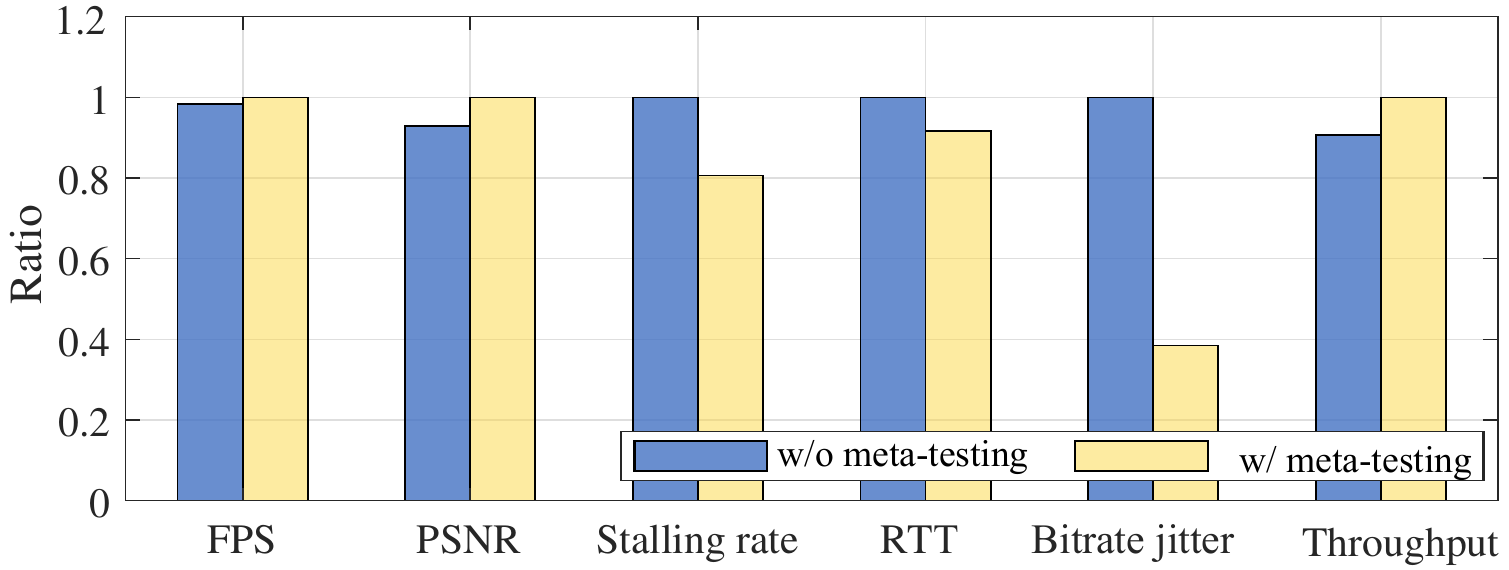}
		\caption{Comparison of Fiammetta w/ and w/o meta-testing.}\vspace{-0.4cm}
		\label{fig:w/wo meta-testing}	
	\end{figure}

	In this subsection, we proceed to investigate whether the gains of Fiammetta are due to the fast adaptation of online meta-testing to changing ``network states'' or to the superiority of the initial NN itself. We conduct experiments to compare the performance of Fiammetta w/ and w/o meta-testing. Through the results in Fig.~\ref{fig:w/wo meta-testing}, we note that Fiammetta w/ meta-testing achieves performance gains across all application- and transport-layer metrics: \textit{(\romannumeral1)} For the application-layer metrics, the most significant improvements come from the 7.5\% increase in PSNR and the 19.5\% reduction in stalling rate. \textit{(\romannumeral2)} For the transport-layer metrics, the throughput is improved by 9.7\%, with reductions in RTT and bitrate jitter of 7.8\% and 60.8\%, respectively, compared to Fiammetta w/o meta-testing.
	%has seen an improvement in performance across all metrics,
 	
 	\vspace{-0.2cm}
 	
 	\section{Related Work}\label{related}
 	\textbf{Interactive video adaptation.} %Recent years have witnessed the video industry as a whole shifting from a majority VoD approach to a live/interactive video approach. %Unlike VoD streaming with multiple seconds of buffer, interactive video QoE imposes the most stringent latency (few hundreds of milliseconds) and increasing quality requirements. 
 	Much effort has gone into developing bitrate adaptation algorithms for interactive video streaming, which falls into two categories: \textit{(\romannumeral1)} rule-based algorithms~\cite{carlucci2016analysis,cardwell2017bbr,varma2015internet}, and \textit{(\romannumeral2)} learning-based algorithms~\cite{zhou2019learning,zhang2020onrl,zhang2021loki,zhang2022deepcc,ma2022multi}. Rule-based algorithms~\cite{carlucci2016analysis,cardwell2017bbr,varma2015internet} typically follow AIMD-like approaches for bitrate adaptation, based on loss, delay and delay jitter, etc. However, such pre-programmed universal rules can hardly fit diverse and heterogeneous modern networks. The learning-based algorithms consist of offline~\cite{zhou2019learning} and online learning approaches~\cite{zhang2020onrl,zhang2021loki,zhang2022deepcc,ma2022multi}. For instance, Concerto~\cite{zhou2019learning} adopts imitation learning offline. OnRL~\cite{zhang2020onrl} is the first to achieve fully online learning in interactive video applications via federated learning. Loki~\cite{zhang2021loki} proposes tight coupling between RL and GCC. Notwithstanding, the offline-trained model still suffers from fixed parameters. These online-learning  algorithms~\cite{zhang2020onrl,zhang2021loki,zhang2022deepcc,ma2022multi}, however, rely on a large amount of data/time, which hinders fast adaptation to changing ``network states''. Besides, Salsify~\cite{fouladi2018salsify} designs customized codec to address the mismatch between the actual codec bitrate and transport/target bitrate. In addition, many parallel studies~\cite{mao2017neural,zuo2022adaptive,yan2020learning,huang2020quality,akhtar2018oboe} target bitrate adaptation in VoD streaming, which also provide much inspiration for Fiammetta.
	 
 	\textbf{Meta-reinforcement learning.}  %As the name implies, few-shot learning refers to the practice of feeding a learning model with few training data. 
 	Meta-learning~\cite{finn2018learning} is a typical framework for addressing challenging few-shot learning setting by providing reasonable assumptions : \textit{(\romannumeral1)} a task distribution $p(\Gamma)$ exists, \textit{(\romannumeral2)} the target new task obeys $p(\Gamma)$; \textit{(\romannumeral3)} a meta-training set sampled from $p(\Gamma)$ can be obtained to train the model, and \textit{(\romannumeral4)} another meta-testing set sampled from $p(\Gamma)$ can be used to evaluate the expected performance~\cite{finn2018learning}. Fiammetta basically follows these assumptions.  The research on meta-RL includes meta-learning initial NN parameters~\cite{finn2017model,nichol2018reptile}, loss functions~\cite{baik2021meta}, adaptation process~\cite{xu2018meta}, which are all key components of RL. Among them, MAML~\cite{finn2017model} is a landmark work that adjusts simply initial NN parameters, which is sample efficient, lightweight and extremely suitable for few-shot learning settings. In addition, Fiammetta is highly compatible with MAML constraints such as fixed NN structure (e.g., input/output sizes) and strong correlation among tasks.
 	
 	%In what follows, the research on meta-RL falls 
 	\vspace{-0.1cm}

 	 \section{Conclusion}\label{conclusion}
 	 %To our knowledge, 
 	 Fiammetta represents the first meta-RL-based bitrate adaptation algorithm for interactive video system. It marks a new optimization modality that fast adaptation to changing network states becomes a reality by few-shot learning. % with few gradient updates. 
 	 This system is inspired by an observation that network sequences exhibit noticeable short-term continuity sufficient for few-shot learning, drawn from a real-world measurement study on WeChat for Business interactive video system. 
 	 We evaluate Fiammetta on a local testbed with network links following WeChat for Business traces. Experimental results confirm the superiority of Fiammetta compared to state-of-the-art algorithms. %We believe this design can contribute to more video communication applications, such as VoD, 360 panoramic video, virtual reality, etc., by providing new directions for video streaming optimization.
 	 
 	 \section*{Acknowledgement}
 	 This work was supported in part by the National Key R\&D Program of China under Grant 2020YFB1806600, National Science Foundation of China with Grant 62071194, the Key R \& D Program of Hubei Province of China under Grant No. 2021EHB002, Knowledge Innovation Program of Wuhan-Shuguang, Tencent Rhino-Bird Focus Research Project of Basic Platform Technology 2021, the European Union's Horizon 2020 research and innovation programme under the Marie Skłodowska-Curie grant agreement No 101022280.
 	 
 	 %the MUMIMO communication in mobile scenarios by providing new
 	 %insights on channel prediction and link adaptation.

 	 %We believe OnRL hints on a new direction that embraces online learning into more video communication applications, such as VoD, 360 panoramic video, virtual reality, etc.

%\balance

\bibliographystyle{unsrt}
\bibliography{IEEEabrv,./Fiammetta}

\end{document}